\documentclass{ptephy_v1}

\preprintnumber{XXXX-XXXX} 
\usepackage{here}

\newcommand{\average}[1]{\ensuremath{\langle #1 \rangle} }

\begin{document}
\title{Developing an end-to-end simulation framework of supernova neutrino detection}


\author{Masamitsu Mori}
\affil{Department of Physics, Faculty of Science, Kyoto University, Kyoto 606-8502, Japan \email{mori.masamitsu.33s@st.kyoto-u.ac.jp}}

\author[2,3]{Yudai Suwa}
\affil{Department of Astrophysics and Atmospheric Sciences, Kyoto Sangyo University, Kyoto 603-8555, Japan}
\affil[3]{Center for Gravitational Physics, Yukawa Institute for Theoretical Physics, Kyoto University, Kyoto 606-8502, Japan}

\author[4]{Ken'ichiro Nakazato}
\affil[4]{Faculty of Arts and Science, Kyushu University, Fukuoka 819-0395, Japan}

\author[5]{Kohsuke Sumiyoshi}
\affil[5]{National Institute of Technology, Numazu College, Shizuoka 410-8501, Japan}

\author[6]{Masayuki Harada}
\affil[6]{Department of Physics, Okayama University, Okayama 700-8530, Japan}

\author[7]{Akira Harada}
\affil[7]{Institute for Cosmic Ray Research, University of Tokyo, Kashiwa 277-8582, Japan}

\author[6,8]{Yusuke Koshio}
\affil[8]{Kavli Institute for the Physics and Mathematics of the Universe (Kavli IPMU, WPI), Todai Institutes for Advanced Study, \\ University of Tokyo, Kashiwa 277-8583, Japan}

\author[1,8]{Roger A. Wendell}

%

\begin{abstract}%

Massive stars can explode as supernovae at the end of their life cycle, releasing  neutrinos whose total energy reaches $10^{53}{\rm\ erg}$. Moreover, neutrinos play key roles in supernovae, heating and reviving the shock wave as well as cooling the resulting proto-neutron star. Therefore, neutrino detectors are waiting to observe the next galactic supernova and several theoretical simulations of supernova neutrinos are underway. While these simulation concentrate mainly on only the first one second after the supernova bounce, the only observation of a supernova with neutrinos, SN 1987A, revealed that neutrino emission lasts for more than 10 seconds. For this reason, long-time simulation and analysis tools are needed to compare theories with the next observation. Our study is to develop an integrated supernova analysis framework to prepare an analysis pipeline for treating galactic supernovae observations in the near future. This framework deals with the core-collapse, bounce and proto-neutron star cooling processes, as well as with neutrino detection on earth in a consistent manner. 
 We have developed a new long-time supernova simulation in one dimension that explodes successfully and computes the neutrino emission for up to 20 seconds. Using this model we estimate the resulting neutrino signal in the Super-Kamiokande detector to be about 1,800 events for an explosion at 10 kpc and discuss its implications in this paper. 
 We compare this result with the SN 1987A observation to test its reliability. 
 
\end{abstract}

\subjectindex{E14, E26, F22}

\maketitle
\section{Introduction}
\label{Introduction}
A massive star leads to a spectacular explosion called a supernova (SN) at the end of its life. The supernova releases $10^{51}{\rm\ erg}$ in the form of ejected matter.  It also emits $10^{53}{\rm\ erg}$, which amounts to 10$\%$ of the solar rest mass energy from gravitational binding energy, in the form of neutrinos within just a few tens of seconds.  Supernova explosions are among the most energetic phenomena in the universe, releasing into space both elements that were synthesized over the entire life of the parent star as well as heavy elements synthesized during the explosion itself. For this reason, supernovae provide important information about the chemical evolution of the universe. 

In order to study the supernova explosion mechanism itself, it is important to study the neutrinos emitted during the evolution of core-collapse supernovae. The standard scenario of core-collapse supernovae is believed to be the so-called neutrino-driven explosion, in which the interaction of neutrinos and matter plays an essential role~\cite{Kotake_2012,Mueller_2016, Janka_2017, hori18}. Neutrinos are involved in all phases of the explosion through trapping, emission and absorption in the collapse, as well as during the bounce and explosion resulting in the formation of a neutron star. This general scenario was shown to be viable by the detection of neutrinos from SN 1987A \cite{Hirata:1987hu,Bionta_1987,Alexeyev_1988}.  However, the detailed mechanism has not been clarified because only about 20 neutrino events were observed from SN 1987A, which was only able to provide information on the scale of the total energy budget and pointed to neutrino diffusion over long time scales from a compact object.
Since that time neutrino detectors have been waiting to detect the next galactic supernova.
For example, Super-Kamiokande~\cite{FUKUDA2003418}, which is a large water Cherenkov detector, is expected to detect thousands of events from a supernova occurring in our Galaxy. 
Once such neutrinos are detected with high statistics, they will provide valuable information that can elucidate the full dynamics of core-collapse supernovae and the subsequent formation of neutron stars.

It is therefore important to prepare various sets of theoretical models and their predicted signals to allow for rapid and thorough exploration of the possible  physical scenarios using any observed supernova neutrinos. Studies of supernova neutrinos covering their evolution from the explosion to the neutron star birth have so far been limited to a few explosion models. Although there are many numerical studies of supernovae, most simulations aim to understand the supernova mechanism focusing on only the first second of the process, which is critical to determine whether a supernova explodes at all. Accordingly, there are only a few studies of future supernova neutrino detection and they typically adopt a particular model which successfully explodes and forms a neutron star~\cite{totani, hued10,fisc10,Roberts:2012um,MartinezPinedo:2012rb,Fischer:2018kdt,Bollig:2020phc}. There are also other studies on the number of expected events based on simulations with prescribed explosions \cite{Abe_2016,Suwa_2019} as well as approximate analytic solutions~\cite{suwa20}. Without sufficient simulations covering the entire evolution of the supernova explosion, neutron star formation, and the corresponding neutrino event prediction in a terrestrial detector, it is difficult to compare theoretical predictions with a real observation to obtain constraints on model parameters. 

\begin{figure}
  \centering
  \includegraphics[width=10cm]{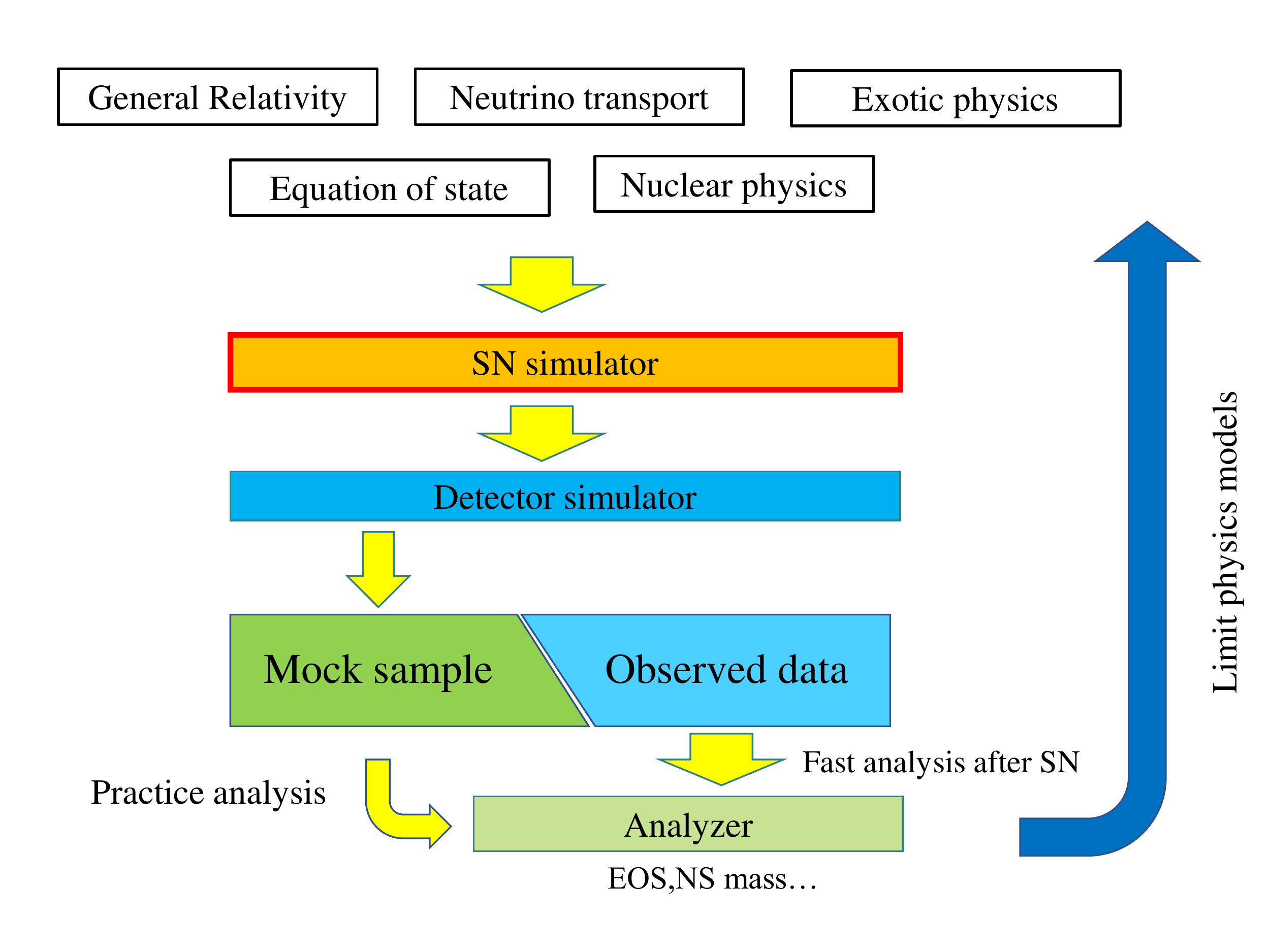}
  \caption{Schematic diagram of the integrated supernova framework. This framework comprises a long-time supernova simulator, a detector simulator and an event analyzer. Before detection of a supernova, we provide mock samples to allow for practice analysis. After detection, we can quickly analyze the supernova to provide feedback to various supernova models.}
  \label{figure_analyzer}
\end{figure}

To overcome these difficulties, recent supernova studies predict neutrino signals in terrestrial detectors~\cite{Li:2020ujl,Warren:2019lgb,Nagakura:2020bbw,Suwa_2019} and to the same end we 
are developing a framework for supernova-neutrino analysis (Fig.~\ref{figure_analyzer}).
Our framework consists of a supernova simulator, a detector simulator and a supernova analyzer. The supernova simulator uses a package of simulation software to provide theoretical predictions of the neutrino emission over long time scales, covering the core-collapse, bounce, explosion, and cooling phase of the proto-neutron star (PNS). The detector simulator provides mock data representing detected neutrino events using the output of the supernova simulator. The analyzer aims to provide methods to compare the mock sample and the observational data to quickly analyze the properties of a supernova based on the results from the supernova and detector simulators. The purpose of our framework is to connect the numerical simulations of the stellar collapse and explosion with the prediction and observation of neutrino events at detectors on earth. We want to bridge the gap 
from the simulation to the detector so that one can analyze a burst in a systematic manner and to  eventually extract model information from the supernova explosion. 

The first steps of the framework development are the creation of a supernova simulator that performs long-time simulations of over 20 seconds and a detector simulator that generates mock data by Monte-Carlo sampling from the output of the supernova simulator.  
We provide the supernova analyzer to make a rapid analysis of a real supernova burst and to find connections between such an observation and theory using the evolution of the number of events and their energy as a function of time. 
Note that previous studies have also attempted to produce the late time neutrino spectra~\cite{Nagakura:2020bbw}. In the present work we additionally discuss the mean and variance of the number of events and their energies in order to provide robust information even when statistics are small, which may be beneficial immediately following a supernova observation.


In this paper, we demonstrate the flow of our framework by showing results of the supernova simulation described below and mock data generated therewith. We describe the new long-time simulation in \S\ref{sec:method} and \S\ref{results}, which correspond to the ``SN simulator'' in Fig.~\ref{figure_analyzer}. We show examples of the construction of mock data sampled from our simulation and analyses of event rates at Super-Kamiokande in \S\ref{analysis_demo}, corresponding to the ``Detector simulator'' and ``Analyzer'' in Fig. \ref{figure_analyzer}, respectively.

\section{Hydrodynamics simulation Methods}\label{sec:method}
\label{Method}

We employ GR1D \cite{O_Connor_2010, O_Connor_2015} for hydrodynamic simulations in which general relativistic hydrodynamics equations, as well as multi-energy neutrino radiation transport equations, are solved in a spherically symmetric geometry.\footnote{The code is publicly available at https://www.GR1Dcode.org.} Neutrino transport is solved using the truncated momentum formalism of Ref.~\cite{Shibata:2011kx}.

Since GR1D is open source it has the advantage of allowing the community to reproduce and extend our results, which is not necessarily the case with other custom codes.\footnote{The truncated momentum formalism used in GR1D is likely to produce qualitatively reasonable results and we expect that our conclusions in this work will not change significantly even if we employ more sophisticated numerical schemes for computing the details of neutrino transfer. In particular GR1D addresses three flavor neutrino transport while recently calculations using six flavors have also been performed~\cite{PhysRevLett.119.242702,Fischer:2020vie}. The additional flavors are likely to influence the initial collapse and may also affect the late stage neutrino spectra and will therefore be considered in a future work.} 
We employ the nuclear equation of state (EOS) based on the density-dependent relativistic mean-field (DD2) model \cite{Hempel:2009mc}. Neutrino interactions are calculated using a numerical table made with NuLib\footnote{The code is publicly available at https://www.nulib.org.} \cite{O_Connor_2015}.
We employ a 9.6 $M_\odot$ zero-metallicity progenitor provided by A. Heger (2016, private communication), which has been used in the previous works and shown to explode even assuming spherical symmetry \cite{Melson_2015, Radice_2017}. This allows us to perform a long-time simulation from the initial core-collapse through the PNS cooling without introducing other phenomenological modeling.

Multi-dimensional effects such as the standing accretion-shock instability~\cite{Nagakura:2020qhb,Walk:2018gaw}, rotation~\cite{Shibagaki:2020ksk} and convection~\cite{Roberts_2012} also play important roles in supernovae and are expected to influence the neutrino signal somewhat. 
However in this study, we do not consider multi-dimensional effects due to their computational expense. 
Indeed, in the case of a light progenitor around 9.0 $M_\odot$, multi-dimensional effects are expected to have little impact on the neutrino signal~\cite{Nagakura:2020qhb} so we ignore them here. Note that there are methods under development that approximately incorporate multi-dimensional effects into spherically symmetric simulations~\cite{Warren:2019lgb,Couch_2020}.

In the following we explain our modifications of GR1D for long-term simulations in \S\ref{sec:modification}, the grid settings in \S\ref{grid_setting}, and the neutrino reactions in \S\ref{neutrino_reaction_table}.

\subsection{Modifications for long-term simulations}\label{sec:modification}

GR1D is publicly available software for the simulation of core-collapse supernova explosions that is well-suited to modelling the accretion phase, which typically occurs up to one second after the bounce. 
Since we are interested in the PNS cooling phase in this paper, which happens later, we have modified the original GR1D in two ways.
First, the original grid profiles of GR1D are optimized to resolve the supernova shock evolution, which is not suitable for the PNS cooling phase. To perform the PNS cooling simulation we have adjusted the grids so as to resolve the steep density gradient at the surface of the PNS. Details are presented in the next section. We also change the CFL number from the original setting of 0.5 to 0.25 starting 8~s after the bounce. 
Second, we find that the long-term evolution causes some thermodynamic quantities such as the density, temperature, electron fraction, and $\eta=\mu/k_BT$, with $\mu$ being the chemical potential of electrons, to exceed the bounds of GR1D's numerical tables.
To avoid such overflows, we have reproduced and extended those tables to cover a wider range of parameters.
Further, we have chosen to fix any values that exceed the limits of the new tables to their closest extremum. 
Details are given in \S\ref{neutrino_reaction_table}.

\subsection{Grid settings}\label{grid_setting}

\begin{figure}
    \centering
    \includegraphics[width=10cm]{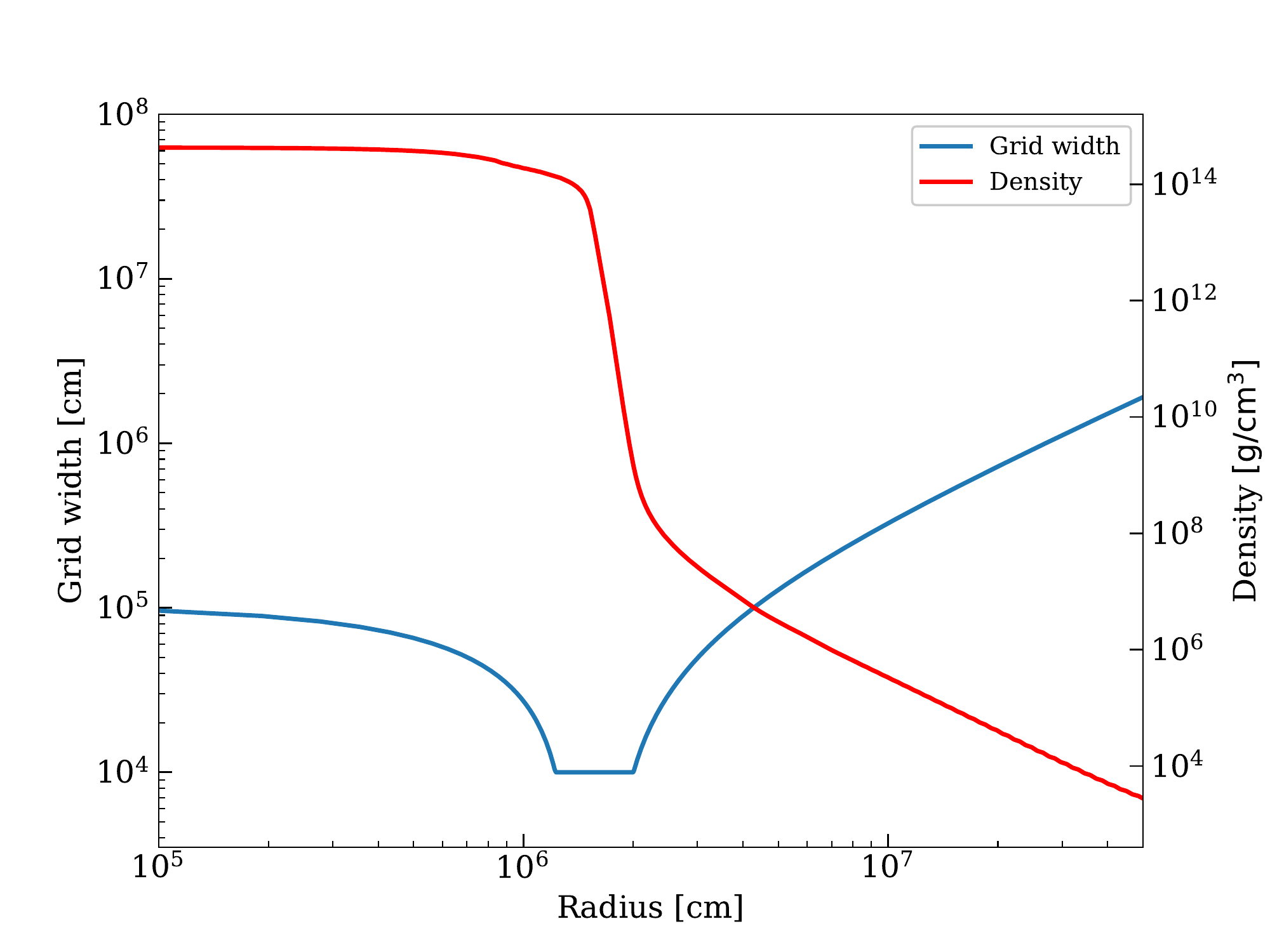}
    \caption{Grid width and density profile. 
    The blue line shows the grid width and the red line shows the density profile three seconds after the bounce.}
    \label{fig:grid_width}
\end{figure}

During PNS cooling, the surface density gradient is extremely steep, decreasing from $\sim 10^{14}$ g cm$^{-3}$ to $\sim10^{8}$ g cm$^{-3}$ within 20 km (see the red line in Fig. \ref{fig:grid_width}). 
Therefore we have employed fine grids to resolve this gradient. 
For this purpose we use {\tt custom2}, which allows the 30 innermost zones to have zone widths that decrease logarithmically from 1 km to 0.1 km and allows the intermediate 78 zones to have a constant width (0.1 km) with an otherwise logarithmic progression that extends up to 5000 km (see the blue line in Fig. \ref{fig:grid_width}). 
There are 300 total grid points used throughout the simulation. 
As shown in Fig. \ref{fig:grid_width} the finest resolution grid is assigned for the steepest density gradient at the PNS surface. 

\subsection{Neutrino reaction table}\label{neutrino_reaction_table}

In GR1D the neutrino reactions are given by a numerical table that is computed in advance using NuLib and DD2 as an EOS input.
Since the original table does not cover the whole thermodynamic range necessary for the PNS cooling phase, we have produced an expanded table. 
It spans $\rho=10^{6-15.5}$ g cm$^{-3}$ with 82 logarithmically-spaced points  $T=0.05-150$ MeV with 65 logarithmically-spaced points, $Y_e=0.015-0.55$ with 82 linearly-spaced points, and $\eta=0.1-100$ with 61 logarithmically-spaced points.

Neutrino interactions with matter are an integral part of the explosion's evolution.
In this work we consider the emission and absorption reactions
\begin{align}
    \nu_{\rm e} + n &\leftrightarrow p + e^{-}, \\
    \bar{\nu}_{\rm e} + p &\leftrightarrow n + e^{+}, \\
    \nu_{\rm e} + (A,Z) &\leftrightarrow (A,Z+1) + e^{-},
\end{align}
where $\nu_{\rm e}$, $\bar{\nu}_{\rm e}$, $p$, $n$, $e^{-}$, $e^{+}$, and $(A,Z)$ represent electron-type neutrinos, electron-type antineutrinos, protons, neutrons, electrons, positrons, and a nucleus with mass number $A$ and atomic number $Z$, respectively. These reactions are calculated based on \cite{Burrows_2006} with weak-magnetism and recoil corrections from \cite{Horowitz:2001xf} taken into account.
Neutrino absorption on heavy nuclei is also based on \cite{Burrows_2006, Bruenn:1985en}.

Similarly we consider elastic scattering via the following reactions 
\begin{align}
    \nu + \alpha &\rightarrow \nu + \alpha, \\
    \nu_{\rm i} + p &\rightarrow \nu_{\rm i} + p, \\
    \nu_{\rm i} + n &\rightarrow \nu_{\rm i} + n, \\
    \nu + (A,Z) &\rightarrow \nu + (A,Z),
\end{align}
where  $\alpha$ is the helium nucleus, $\nu$ indicates that the reaction is insensitive to the neutrino flavor, 
and $\nu_{\rm i}$ indicates that the reaction depends on flavor. These are also based on \cite{Burrows_2006,Bruenn:1985en}. 
Inelastic scattering 
\begin{equation}
    \nu_{\rm i} + e^- \rightarrow \nu_{\rm i}^{\prime} + {e^{-}}^{\prime},
\end{equation}
has been calculated following~\cite{Bruenn:1985en}.

We consider the following thermal processes,
\begin{align}
    e^{-} + e^{+} &\rightarrow \nu_{\rm x} + \bar{\nu}_{\rm x}, \\
    N + N &\rightarrow N + N + \nu_{\rm i} + \bar{\nu}_{\rm i},
\end{align}
which have been calculated by \cite{Burrows_2006,Bruenn:1985en}.
Note that electron-positron annihilation is considered only for $\nu_{\rm x}$ throughout the simulation, where $\nu_{\rm x}$ refers to non-electron-type flavors. 
Including electron flavors results in the simulation stopping prematurely around the bounce. Note though that this reaction has little influence on the late phase neutrino spectra~\cite{O_Connor_2015}. In addition, the rate of electron-positron annihilation is a few orders of magnitude lower than that of nucleon-nucleon bremsstrahlung at energies lower than 50\,MeV~\cite{Kotake:2018ypf}.

Though the original GR1D includes nucleon-nucleon bremsstrahlung for $\nu_{\rm x}$ only, we find that without including 
electron-type neutrinos the average energies of these flavors remain constant at late times, though they are expected to 
decrease based on physical considerations. 
Accordingly, we have added them to the bremsstrahlung process to resolve this issue.
Note that we do not consider medium suppression for simplicity \cite{Kotake:2018ypf}.

\section{Hydrodynamic simulation results}
\label{results}

In this section we present results from our simulation, summarizing the dynamical features in section \ref{dynamical_properties} before 
discussing neutrino luminosities and average energies in section \ref{neutrino_properties}.

\subsection{Dynamical properties}\label{dynamical_properties}
\label{dynamical_properties}
\begin{figure}[htbp]
  \centering
  \includegraphics[width=10cm]{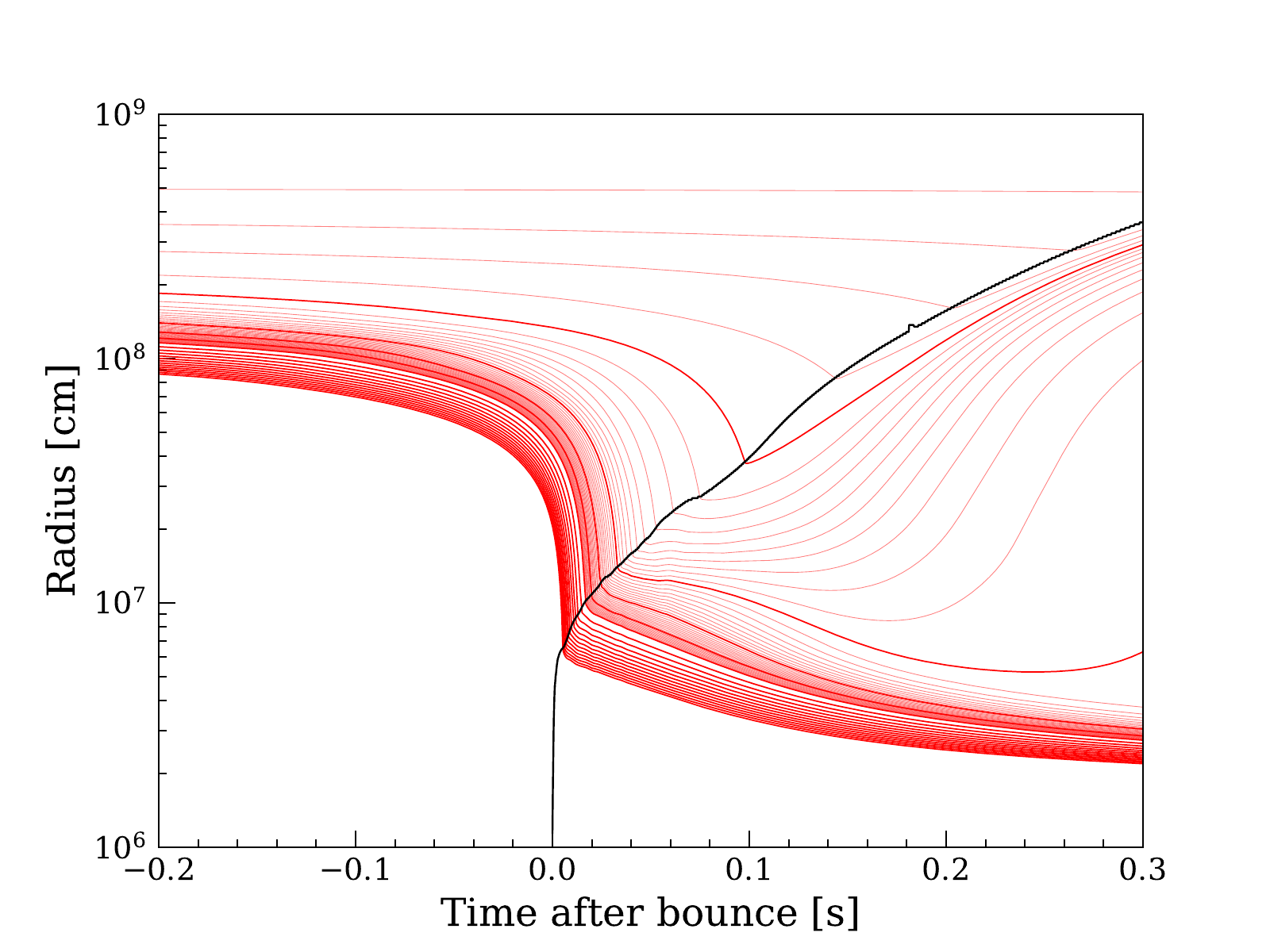}
  \caption{Time evolution of mass shells (red) and the shock wave (black)\label{masscoordinate.pdf}. The thick and thin lines indicate mass shells ranging from $1.20\,M_\odot$ to $1.37\,M_\odot$ at intervals of $10^{-2}\,M_\odot$ and from $1.330\,M_\odot$ to $1.374\,M_\odot$ at intervals of $10^{-3}\,M_\odot$, respectively.}
\end{figure}

Figure~\ref{masscoordinate.pdf} shows the trajectories of several mass shells and the shock wave in red and black, respectively. The shock trajectory clearly shows a successful shock propagation. The innermost red line corresponds to the mass coordinate $1.20\,M_\odot$, while the outermost thick line indicates that of $1.37\,M_\odot$. Here the thick red lines are separated by $10^{-2}\,M_\odot$. In order to see the mass shells outside the PNS in detail, thin red lines for $1.330\,M_\odot$ to up to $1.374\,M_\odot$ are shown in intervals of $10^{-3}\,M_\odot$. At the beginning of the simulation the mass shells fall into the center gradually and then rapidly just before the core bounce. The initial central density of the progenitor is $10^9\,{\rm g\,cm^{-3}}$, and the core collapse continues until the central density exceeds the nuclear saturation density $\sim 10^{14}\,{\rm g\,cm^{-3}}$. The core bounce of the z9.6 progenitor occurs $0.254\,{\rm s}$ after the start of the simulation.

After the core bounce, the mass shells inside $1.36\,M_\odot$ accrete onto the PNS surface, while others are ejected after the passage of the shock. Hence the baryonic mass of the PNS remnant of this model is $1.36\,M_\odot$. 
The shock wave continues to propagate outward until it reaches the simulation boundary of $5000\,{\rm km}$ $0.35\,{\rm s}$ after the core bounce.


Here we define the commonly used {\it diagnostic explosion energy} \cite{Suwa:2009py} as
\begin{equation}
    E_{\rm exp} =\int_{\varepsilon_{\rm bind}>0}{{\varepsilon_{\rm bind}}}dV,
    \label{definition_exp_ene}
\end{equation}
where
\begin{equation}
    \varepsilon_{\rm bind} = \alpha(\rho(c^2+\epsilon+P/\rho)W^2-P)-\rho Wc^2
\end{equation}
is the binding energy, $\alpha = \sqrt{-g_{00}}$ is the lapse function (see O'Connor et al. (2010) \cite{O_Connor_2010} for details), $\rho$ is the rest-mass density, $\epsilon$ is the specific internal energy, $P$ is the pressure, $W$ is the Lorentz factor, and $dV$ is the three-volume element for the curved space-time metric. 
With this definition the diagnostic explosion energy of this model converges to $4.2\times 10^{49}\,{\rm erg}$, which is significantly smaller than the observed value of $10^{51}\,{\rm erg}$ \cite{Hamuy:2002qx}.
However, this value is consistent with that of previous studies that used the same progenitor model \cite{Melson_2015,Radice_2017}. 
Observations of SNe indicate that they have a rather broad distribution of  explosion energies, from $\sim 10^{50}$ erg to $10^{52}$ erg \cite{Muller:2017bdf}.
The explosion energy is naively determined using the binding energy of progenitor layers in the vicinity of the final remnant mass of the compact object so that stars with small binding energies are possible candidates for weak explosions (see e.g., \cite{Suwa:2015saa}).


\begin{figure}[htbp]
  \centering
  \includegraphics[width=10cm]{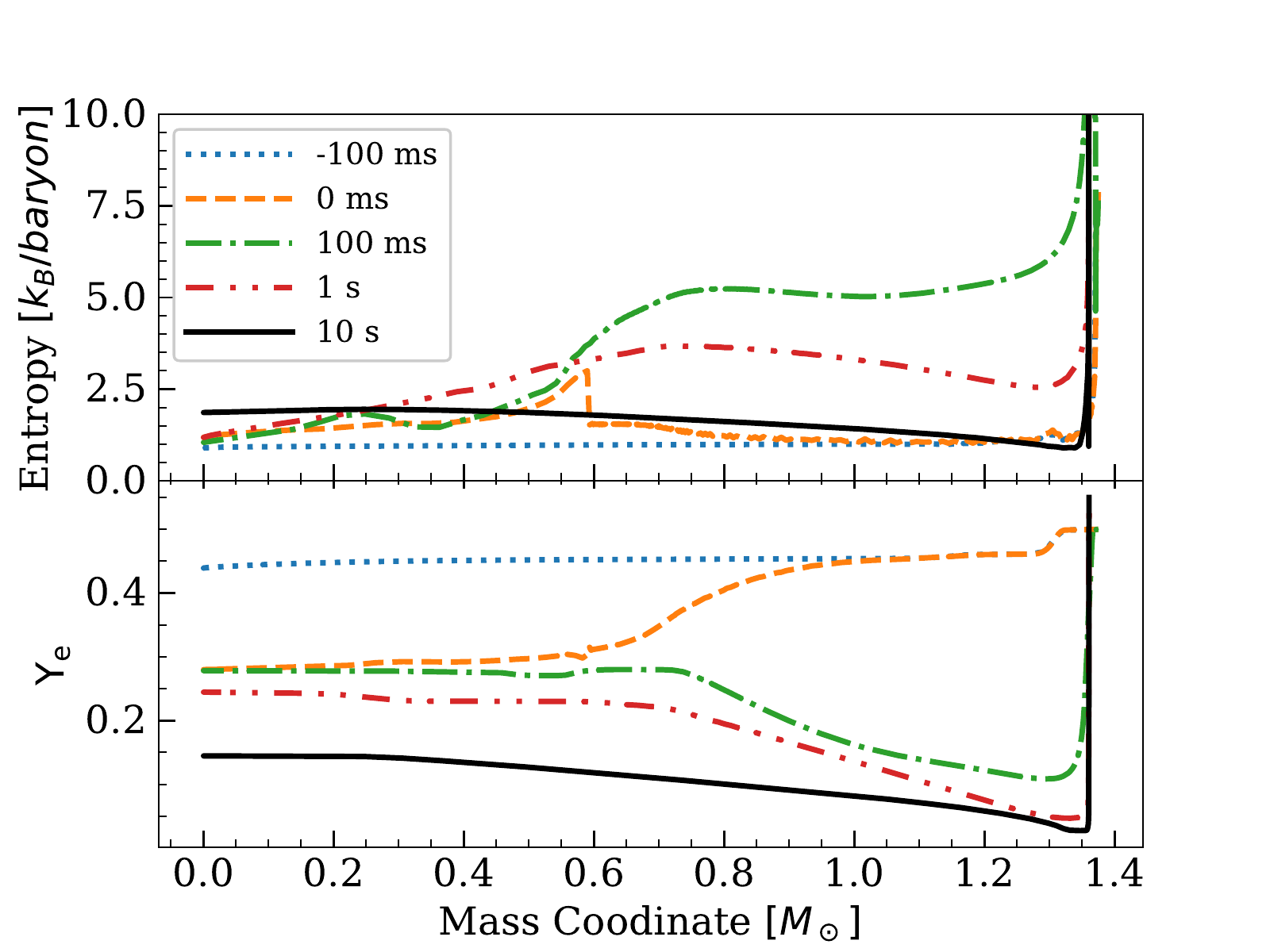}
  \caption{\label{entropy_ye_times} Radial profiles of the entropy and the electron fraction. The horizontal axis is the baryonic mass coordinate. The time in the legend is defined relative to the core bounce: positive values mean after the bounce and the negative values mean before. The upper panel shows the entropy, while the lower panel shows the electron fraction. There is a jump at $0.6\,M_\odot$ in the entropy profile and a small dip in the electron fraction profile due to the shock wave at the core bounce.}
\end{figure}

Radial profiles of the entropy and $Y_{\rm e}$ at different times are shown in Fig. \ref{entropy_ye_times}. 
Before the bounce at $-100\,{\rm ms}$ (blue dotted line), the entropy is as low as $1.2\,k_{\rm B}/{\rm baryon}$ and is constant with respect to the mass coordinates. The electron fraction at that time is constant and as high as $0.45$ up to $1.3\,M_\odot$ and 
slightly increases to $0.5$ at $1.4\,M_\odot$. 
At the time of the core bounce (orange dashed line), the shock wave appears at the mass coordinate $0.6\,M_\odot$ 
resulting in a sudden jump in entropy from $1.5\,k_{\rm B}/{\rm baryon}$ to $3.0\,k_{\rm B}/{\rm baryon}$, 
while the entropy at external coordinates is nearly constant. 
The electron fraction $Y_{\rm e}$ is $\sim 0.3$ up to the mass coordinate $0.6\,M_\odot$ and increases to $0.45$ in the outer regions. 
After the core bounce (green dash-dotted and red dash-double-dotted lines), the entropy increases up to $\sim 0.8\,M_\odot$, 
becomes almost constant (at $100\,{\rm ms}$) or decreases (at $1\,{\rm s}$) outside, and suddenly increases from $\sim 1.35\,M_\odot$. 
The electron fraction decreases with the mass coordinate and reaches its minimum value at $\sim 1.35\,M_\odot$. 
At $10\,{\rm s}$ after the core bounce, the entropy takes on an almost constant value of $\sim 2.5\,k_{\rm B}/{\rm baryon}$
and the electron fraction is lower than $0.2$. Both suddenly rise at $\sim 1.35\,M_\odot$ again.


Comparing our results with the simulations in Refs. \cite{Melson_2015} and \cite{Radice_2017}, we note that the shock wave of our simulation propagates outward without slowing down, while the shock waves in previous studies slow slightly at radii around  $2\times10^7{\rm\,cm}$ (Ref. \cite{Melson_2015} indicates stronger deceleration than Ref. \cite{Radice_2017}).
The diagnostic explosion energy ($\sim 4.2\times 10^{49}$ erg) observed in our simulation is close to those of previous studies ($\sim 2\times 10^{49}$ erg in Ref. \cite{Melson_2015} and $\sim$ 1--4$\times 10^{49}$ erg in Ref. \cite{Radice_2017} for 1D simulations) and similarly 
our resultant PNS baryonic mass, $1.36 M_\odot$, is also close to that of Ref. \cite{Radice_2017} ($1.38M_\odot$).
Thus, we conclude that overall behaviors of the simulations are consistent with one another. We speculate that small differences  are due to the differences in the adopted microphysics modeling, such as the EOS or neutrino interaction model.

\subsection{Neutrino properties}\label{neutrino_properties}
\begin{figure}[!h]
  \centering
  \includegraphics[width=10cm]{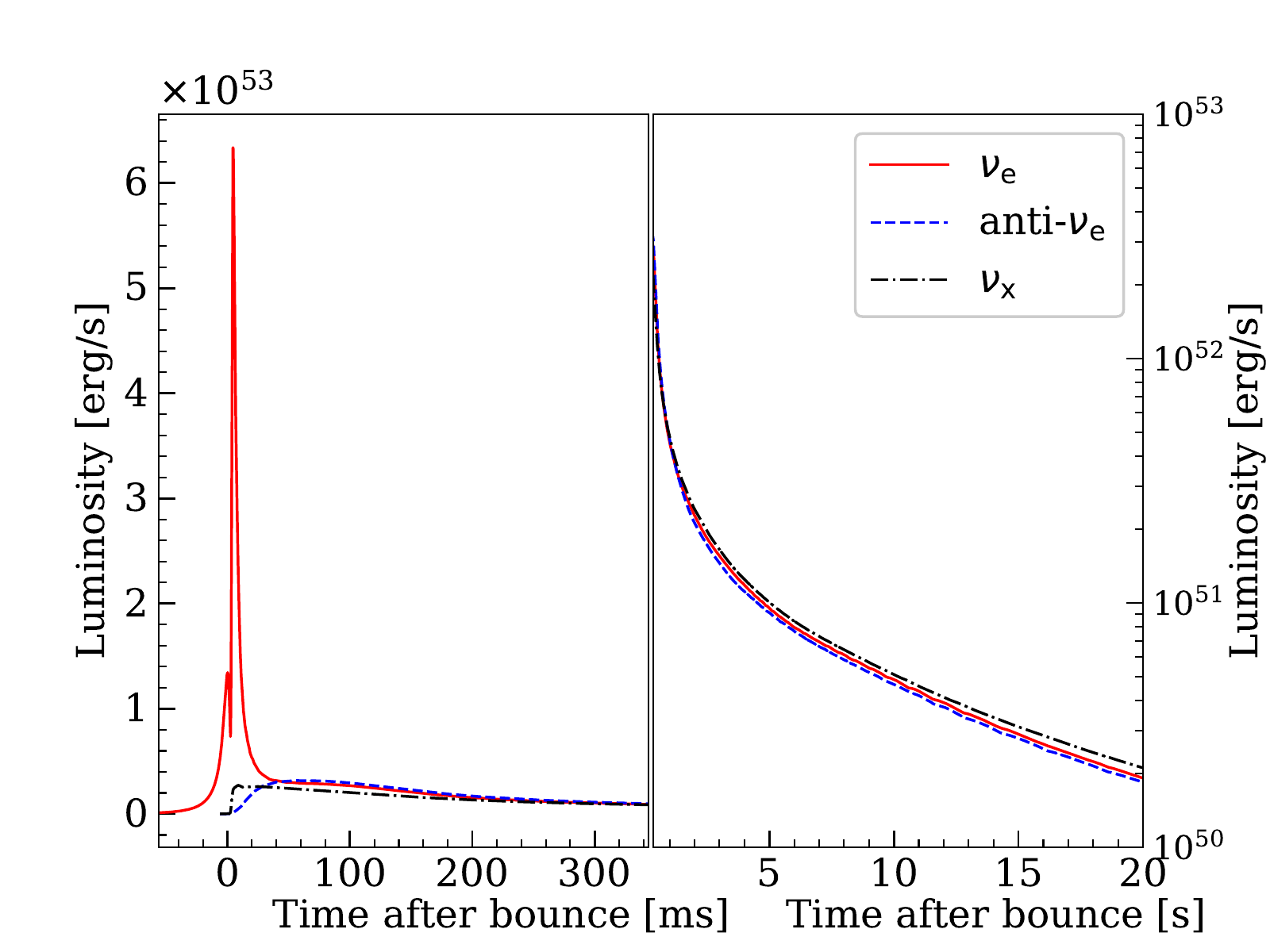}
  \caption{\label{luminosity.pdf} Neutrino luminosities. The left panel shows the early phases, including the core bounce, and the right panel shows later phases. The red, blue, and black lines are the electron neutrino, the anti-electron neutrino, and non-electron-type neutrinos, respectively.}
  \end{figure}

\begin{figure}[!h]
  \centering
  \includegraphics[width=10cm]{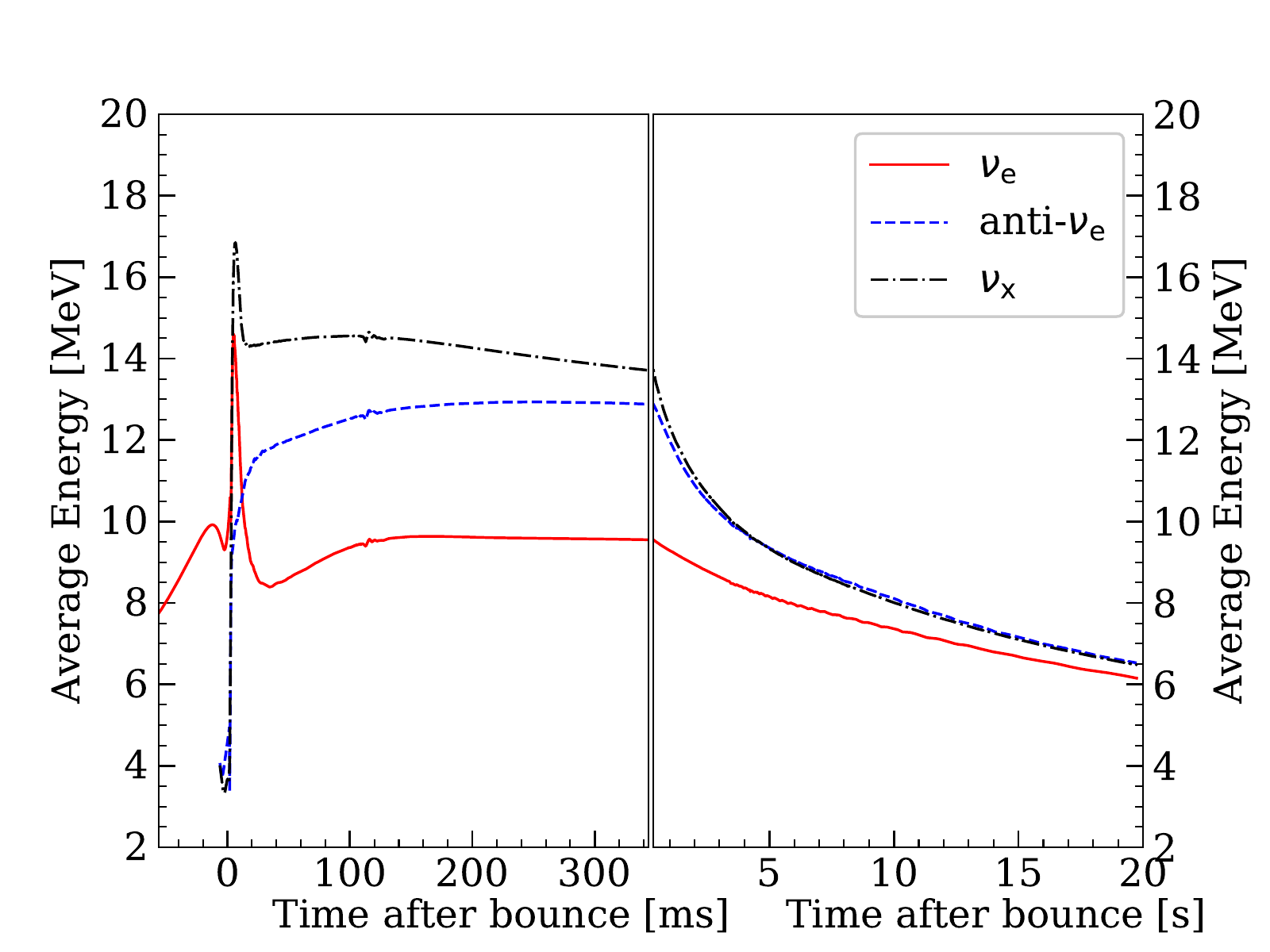}
  \caption{\label{average_energy.pdf} The same as Fig. \ref{luminosity.pdf}, except that the neutrino average energies are displayed.}
\end{figure}

\begin{figure}[!h]
  \centering
  \includegraphics[width=10cm]{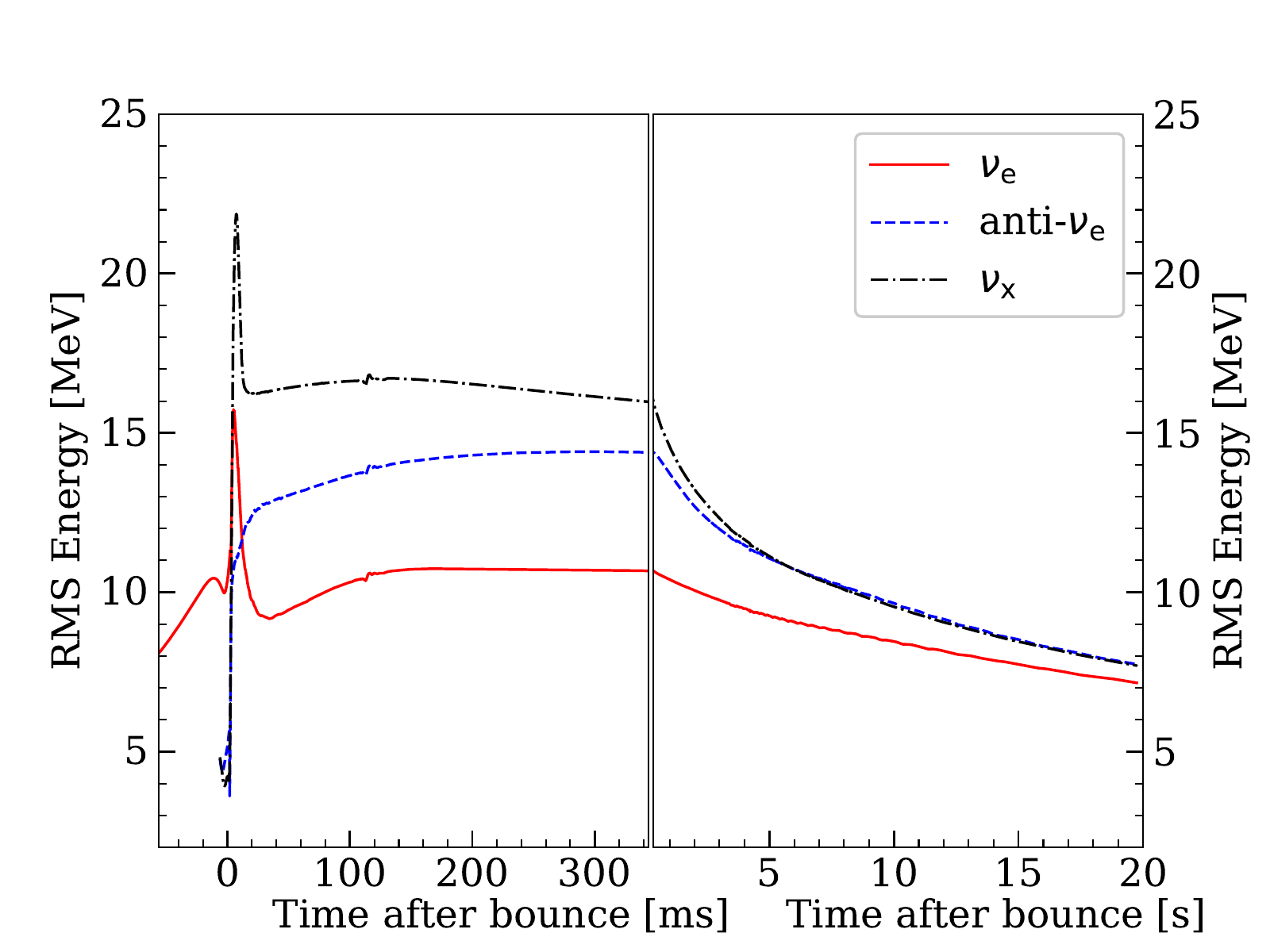}
  \caption{\label{rms_energy.pdf} The same as Figs. \ref{luminosity.pdf} and \ref{average_energy.pdf}, except that the neutrino RMS energies are shown.}
\end{figure}

We simulated the time evolution of neutrinos in the z9.6 model. The luminosity $L_\nu$, average energy $\average{E_\nu}$, and the root-mean-square (RMS) energy $\sqrt{\average{E_\nu^2}}$ are shown in Figs. \ref{luminosity.pdf}, \ref{average_energy.pdf}, and \ref{rms_energy.pdf}, respectively. In this paper, the definitions of these energies are
\begin{align}
    \average{E_\nu} & = \frac{\displaystyle\int E \dfrac{dN}{dE} dE }{ \displaystyle\int \dfrac{dN}{dE} dE }, \\
    \average{E^2_\nu} &= \frac{\displaystyle\int E^2 \dfrac{dN}{dE} dE}{\displaystyle\int \dfrac{dN}{dE} dE},
\end{align}
where $dN(E)/dE$ is the neutrino number density per unit energy.


In Fig. \ref{luminosity.pdf}, the electron neutrino ($\nu_{\rm e}$) luminosity begins to rise at $\sim 20\,{\rm ms}$ before the core bounce, and then falls off temporarily around the bounce due to neutrino trapping. Note that there are few neutrinos of the other types before the core bounce since the electrons are degenerate at the lower temperatures at this stage in the evolution. 
During this period the dominant reaction in detectors on the ground would therefore be electron scattering, which is sensitive to the direction of the incoming neutrino and hence the SN location on the sky.
Shortly after the core bounce, the $\nu_{\rm e}$ luminosity rapidly increases to $6.5\times 10^{53}\,{\rm erg\,s^{-1}}$ due to the neutronization burst and then drops to $\sim 1.0\times 10^{53}\,{\rm erg\,s^{-1}}$. 
Note that the peak of the $\nu_{\rm e}$ luminosity from GR1D is likely to be higher than that of other simulations according to O'Connor et al. (2018)\cite{OConnor:2018sti}. The anti-electron neutrino $\bar{\nu}_{\rm e}$ luminosity gradually increases to the same level as the $\nu_{\rm e}$ luminosity after the core bounce.
For non-electron-type neutrinos, $\nu_{\rm x}$, which represents $\nu_\mu$, $\nu_\tau$, and their anti-particles collectively, the luminosity quickly jumps to $1.2\times 10^{53}\,{\rm erg\,s^{-1}}$. Each of the flavors represented by $\nu_{\rm x}$ has the luminosity shown in the figure. 
After $100\,{\rm ms}$, all luminosities gradually decrease and this trend continues during the PNS cooling phase. 
Finally, all luminosities have nearly the same value at $20\,{\rm s}$.


Figure~\ref{average_energy.pdf} shows that the average $\nu_{\rm e}$ energy is $8\,{\rm MeV}$ initially, increases to $10\,{\rm MeV}$, and then drops slightly during neutrino trapping. At the core bounce the average $\nu_{\rm e}$ energy reaches a peak value of $15\,{\rm MeV}$. 
After the core bounce the average $\nu_{\rm e}$ energy is almost constant around $10\,{\rm MeV}$ and lasts for 
several hundreds of milliseconds. 
The average $\bar{\nu}_{\rm e}$ energy after the core bounce is also constant but higher than that of $\nu_{\rm e}$ because $\bar{\nu}_{\rm e}$ interacts with matter more weakly than $\nu_{\rm e}$ and the $\bar{\nu}_{\rm e}$ neutrinosphere is at a smaller radius than that of $\nu_{\rm e}$.
The average $\nu_{\rm x}$ energy is between $5\,{\rm MeV}$ and $3\,{\rm MeV}$ initially, rises to $17\,{\rm MeV}$ rapidly at the core bounce, and then slightly decreases to $15\,{\rm MeV}$. Afterward, the average energies for all species gradually decrease to $6\,{\rm MeV}$. 
In particular, the average energies of $\bar{\nu}_{\rm e}$ and $\nu_{\rm x}$ energies almost overlap at late times. 
In comparison to the simulation of Ref. \cite{Stockinger:2020hse}, the average energy of $\nu_{\rm e}$ in our simulation is higher by 0.5\,MeV and that of $\nu_{\rm x}$ is lower by 0.5\,MeV in the early phase. These small differences are likely due to differences in the microphysics as mentioned above \S\ref{dynamical_properties}. However these differences are not expected to have a large impact on response of the detector.


As shown in Fig. \ref{rms_energy.pdf}, the behavior of the RMS energies is similar to that of the average energies for all types of neutrinos. 
The RMS energies are related to the pinching parameter of the neutrino spectra.
Both the average and RMS energies are translated to the observed distributions in a terrestrial detector, indicating observations will provide feedback on their initial distributions at the time of the SN.
Note that there is a small ripple at $110\,{\rm ms}$ for all species because the shock wave crosses the point where we calculate the neutrino luminosities, $500\,{\rm km}$, at that time.

\begin{figure}[!h]
  \centering
  \includegraphics[width=10cm]{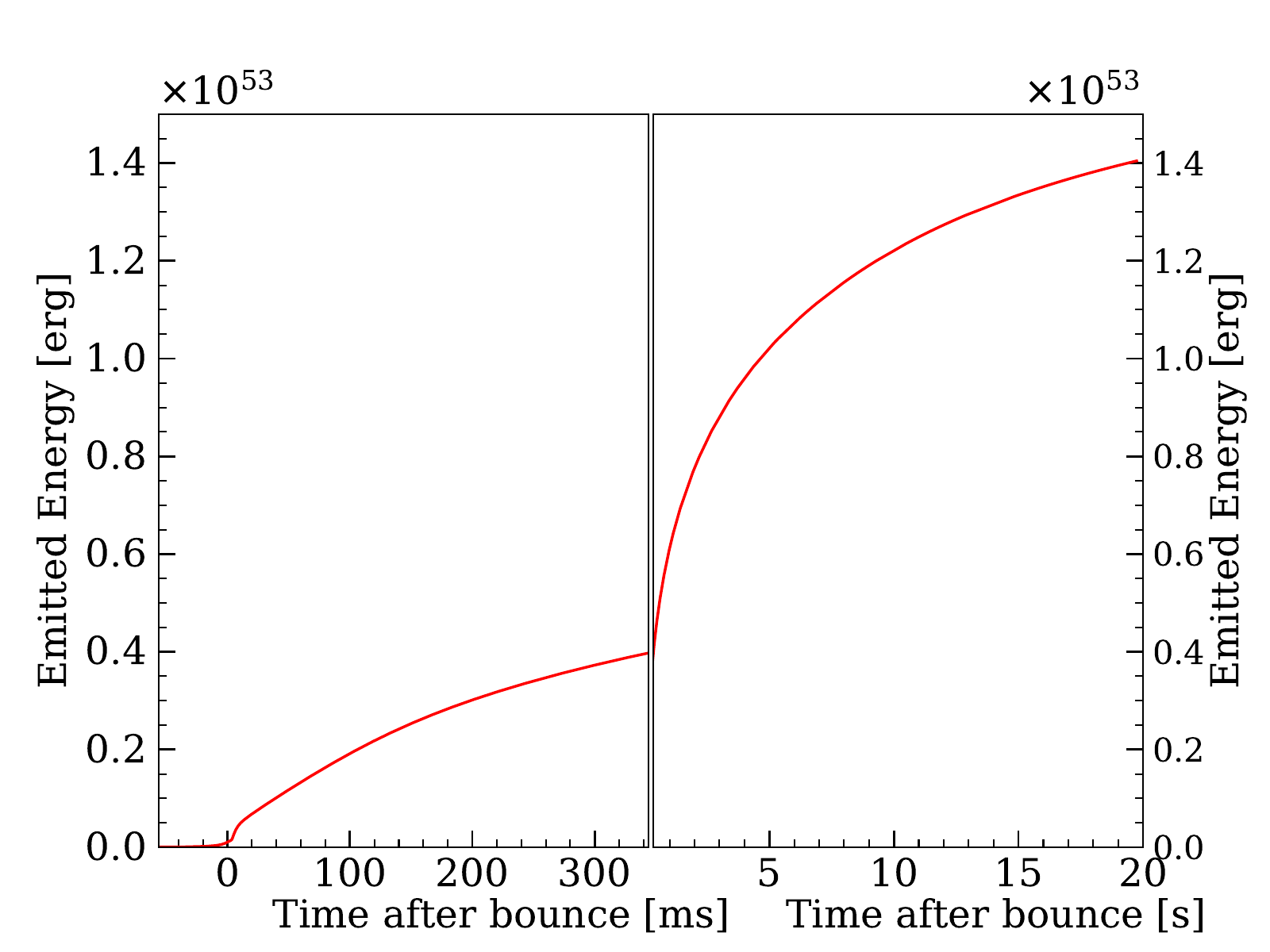}
  \caption{\label{tot_ene} Total energy of emitted neutrinos. The luminosities in Fig. \ref{luminosity.pdf} are integrated and summed  over all flavors up to each point on the horizontal axis.}
\end{figure}

Figure~\ref{tot_ene} shows the total energy of the emitted neutrinos, that is the luminosity in Fig. \ref{luminosity.pdf} has been integrated until the time on the horizontal axis and summed over all flavors. Although the neutrino energy released until $1\,{\rm s}$ is as low as $\sim 0.5\times 10^{53}\,{\rm erg}$, that until $20\,{\rm s}$ is as high as $1.4\times 10^{53}\,{\rm erg}$. 
For this reason it is important to follow the simulation over long times.


\section{Detection simulations}\label{analysis_demo}
In this section we demonstrate the detector simulator part of our analysis framework (Fig. \ref{figure_analyzer}). 
So as to produce mock data samples we perform Monte Carlo simulations of the neutrino detection at Super-Kamiokande (SK) \cite{FUKUDA2003418} based on our numerical model described in the previous sections.

Super-Kamiokande is the largest underground water Cherenkov detector, consisting of a stainless tank whose height is 41.4 m and whose radius is 39.3 m, and is filled with 50~kton of ultra-pure water. 
The tank is separated into two regions, an inner and an outer detector. The inner detector views the water target with 11,129 photomultiplier tubes and the nominal fiducial volume is a 22.5 kton volume therein, which is defined to reduce backgrounds from the detector walls and ensure proper event reconstruction. 
However, for supernova neutrino bursts, many neutrino events will be detected over a time interval short enough that the contribution from background events is expected to be negligible. 
Thus, hereafter we assume the entire inner detector volume of 32.5 kton can be used for observation. 
Since this study is the first step in the construction of our analysis framework, we do not include the details of the detector response for simplicity.
%
Although in the supernova simulations we do not include neutrino oscillations, for the detector simulation we modify the fluxes to  account for neutrino flavor oscillations in vacuum and in matter. 
Following Refs.~\cite{Dighe:1999bi,Wu:2014kaa}, we mix neutrinos as, 
\begin{align}
    F^\prime_{\nu_{\rm e}} &= F_{\nu_{\rm x}},\\
    F^\prime_{\bar{\nu}_{\rm e}} &= pF_{\bar{\nu}_{\rm e}} + (1-p)F_{\nu_{\rm x}},\\
    4F^\prime_{\nu_{\rm x}} &= F_{\nu_{\rm e}} + (1-p)F_{\bar{\nu}_{\rm e}} + (2+p)F_{\nu_{\rm x}},
\end{align}
for the normal hierarchy and 
\begin{align}
    F^\prime_{\nu_{\rm e}} &= (1-p)F_{\nu_{\rm e}} + pF_{\nu_{\rm x}},\\
    F^\prime_{\bar{\nu}_{\rm e}} &= F_{\nu_{\rm x}},\\
    4F^\prime_{\nu_{\rm x}} &= pF_{\nu_{\rm e}} + F_{\bar{\nu}_{\rm e}} + (3-p)F_{\nu_{\rm x}},
\end{align}
for the inverted hierarchy, where $F^\prime_{\nu}$ and $F_{\nu}$ are neutrino fluxes after and before neutrino oscillations, respectively, and $p$ is $0.69$. 
Note that we employ adiabatic flavor conversion for the following reason. The density of the progenitor in this study decreases rapidly outside the core and the region where the resonant (non-adiabatic) conversion can occur is in the vicinity of the iron core ($\lesssim 10^9$ cm). Although the progenitor density scale height is large enough to satisfy the adiabatic condition, the shock wave is expected to change the density structure and lead to non-adiabatic conversion once it reaches the resonant region. With a typical shock velocity of $\sim 10^9$ cm s$^{-1}$, the shock passes through the resonant regime within a few seconds. Thus, resonant conversion does not change the following discussion, which focuses on later timescales.

\subsection{Detection properties}\label{detection_properties}
Here we focus on the number of events in SK, their angular distribution, and the time evolution of the event rate.
Events are simulated assuming supernova distances of 1~kpc, 5~kpc, 10~kpc and 50~kpc. 
Our Monte Carlo simulations are performed as follows: (i) calculate the expected number of events in each time interval, which are chosen to be 0.01 s ($t < 0.744$ s, where $t$ is the time measured relative to the bounce) and 0.1 s ($t > 0.744$ s), (ii) determine the time of each event in the time interval at random, (iii) evaluate the event spectrum in the time interval and (iv) determine the event energy according to that spectrum for each event. 

We consider the inverse beta decay (IBD) reaction,
\begin{equation}
    \Bar{\nu}_{\rm e} + p \rightarrow {\rm e}^{+} + n,
\label{eq:ibd}
\end{equation}
and the electron scattering (ES) reaction,
\begin{equation}
    \nu + \rm{e}^- \rightarrow \nu + \rm{e}^-.
\end{equation}
The IBD reaction has no sensitivity to the SN direction but has a reaction rate in water that is two orders of magnitude larger than that of electron scattering. 
We adopt the IBD cross section of Strumia and Vissaini \cite{2003PhLB..564...42S}.\footnote{We find that the cross section of Vogel and Beacom (1999) \cite{Vogel_1999} provides 2.6\% fewer total events than that of \cite{2003PhLB..564...42S}.}  
On the other hand, electron scattering has a much smaller reaction rate but the strong correlation of the neutrino and lepton directions 
provides sensitivity to the SN direction. 
The cross section for electron scattering is obtained from field theory as
\begin{align}
        \frac{d\sigma}{d\cos{\theta}} =& 4\frac{m_{\rm e}}{E_{\nu}}\dfrac{\left(1+\dfrac{m_{\rm e}}{E_\nu}\right)^2\cos{\theta}}{\left[\left(1+\dfrac{m_{\rm e}}{E_\nu}\right)^2  - \cos^2{\theta}\right]^2}\frac{d\sigma}{dy}, \\
        \frac{d\sigma}{dy} =& \frac{G_{\rm F}^2m_{\rm e}E_{\nu}}{2\pi}\left[A + B(1-y)^2 - Cy\frac{m_{\rm e}}{E_\nu}\right], \\
        y =& \frac{2\dfrac{m_{\rm e}}{E_\nu}\cos^2{\theta}}{\left(1+\dfrac{m_{\rm e}}{E_\nu}\right)^2  - \cos^2{\theta}},
\end{align}
where the light speed $c$ is 1, $G_{\rm F} = 1.166\times10^{-11}{\rm MeV}^2$ is the Fermi coupling constant, $\theta$ is the angle between the directions of the neutrino and the scattered electron and the constants $A$, $B$ and $C$ are summarized in Table \ref{es_reaction_parameters} and depend upon the neutrino flavor\cite{Suzuki1994}.
Incidentally, we do not include neutrino interactions on oxygen \cite{Nakazato_2018}, which have no direction sensitivity and smaller reaction rates than IBD.
We set the detection threshold energy to be 5 MeV for both positrons and electrons. 
\begin{table}[!h]
    \centering
    \begin{tabular}{cccc}
        \hline
         $\nu$:&$A$&$B$&$C$  \\ \hline
         $\nu_{\rm e}$&$(g_{\rm V}+g_{\rm A}+2)^2$&$(g_{\rm V}-g_{\rm A})^2$& $(g_{\rm V}+1)^2-(g_{\rm A}+1)^2$ \\
         $\bar{\nu}_{\rm e}$&$(g_{\rm V}-g_{\rm A}+2)^2$&$(g_{\rm V}+g_{\rm A}+2)^2$& $(g_{\rm V}+1)^2-(g_{\rm A}+1)^2$ \\
        $\nu_{\mu},\nu_\tau$&$(g_{\rm V}+g_{\rm A})^2$&$(g_{\rm V}-g_{\rm A})^2$&$g_{\rm V}^2-g_{\rm A}^2$ \\
        $\bar{\nu}_{\mu},\bar{\nu}_\tau$&$(g_{\rm V}-g_{\rm A})^2$&$(g_{\rm V}+g_{\rm A})^2$&$g_{\rm V}^2-g_{\rm A}^2$ \\
        \hline
    \end{tabular}
    \caption{Constant parameters for electron scattering. Here $g_{\rm V} = -0.5 + \sin{\theta_{\rm W}}$, where $\theta_{\rm W}$ is the Weinberg angle $\sin^2{\theta_{\rm W}}\approx 0.23$, and $g_{\rm A}=-0.5$ \label{es_reaction_parameters}}
\end{table}

\begin{figure}[H]
    \centering
         \includegraphics[width=10cm]{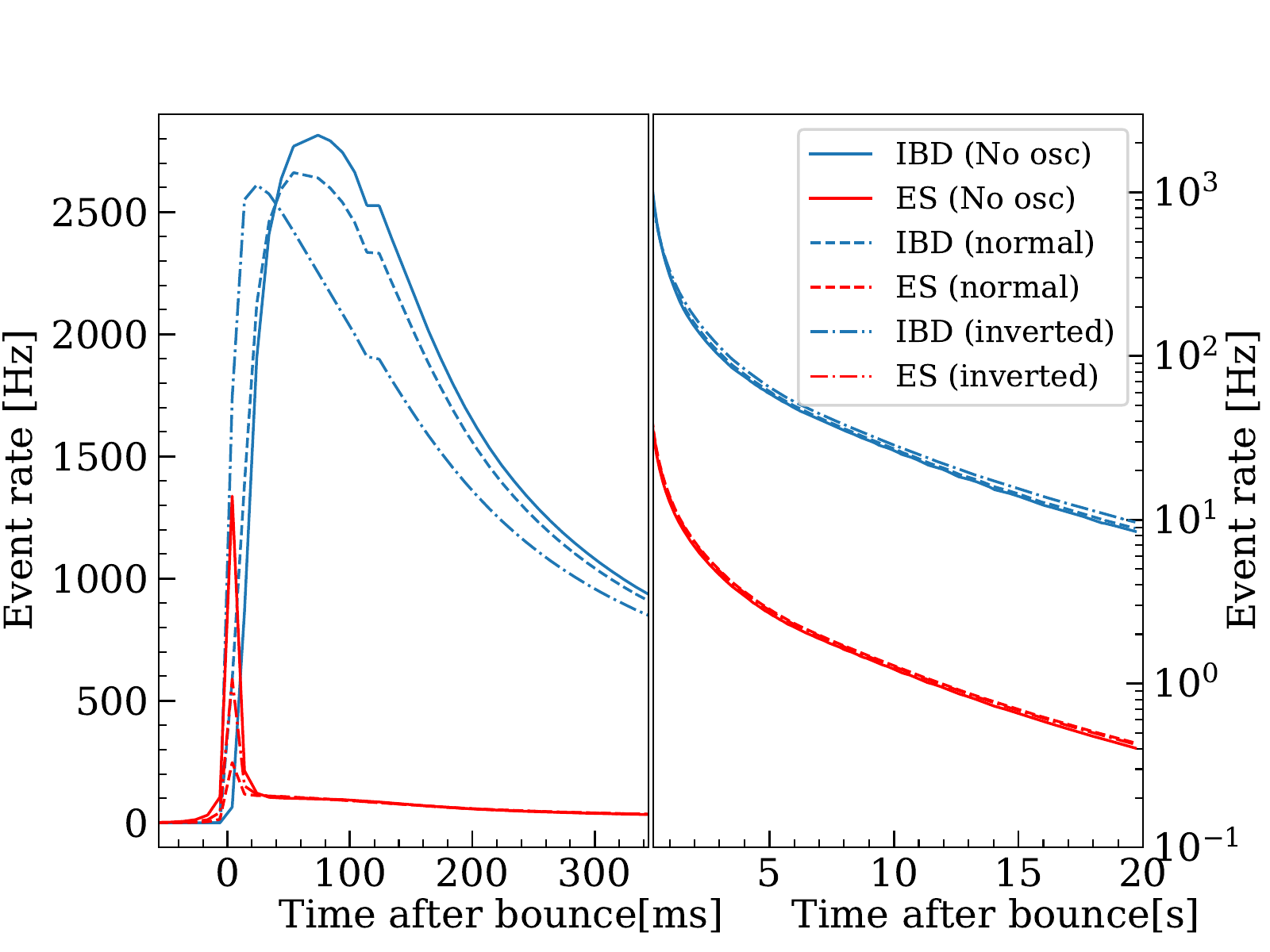}
    \caption{\label{eventrate_ver3} Event rate at SK for a supernova at 10 kpc from the Earth. The solid line is for no oscillation, the dashed line is for normal hierarchy and the dash-dotted line is for inverted hierarchy. The left panel is for the early phase and the right panel is for the late phase.}
\end{figure}

The IBD event rate is shown in Fig. \ref{eventrate_ver3} and the number of IBD events in several time windows is shown in Table \ref{numberofevents} for the case of a supernova at 10~kpc. The total  number of events in our model is of the same order as those of the previous studies~\cite{totani,2013ApJS..205....2N,Abe_2016,Suwa_2019}.
As seen in Table \ref{numberofevents}, half of the events come in the first second, which is a region in which the cooling phase calculation is as important as the early phase calculation.


The IBD rates rapidly increase up to 2800\,Hz for no oscillation, 2700\,Hz for the normal hierarchy and 2600\,Hz for the inverted hierarchy before gradually decreasing to 10\,Hz at 20 seconds (Fig.~\ref{eventrate_ver3}).
Three sharp peaks in the ES rate can be seen around the bounce, which is the only point at which the ES rate dominates the IBD rate for the no oscillations case. Supposing neutrino oscillations, these peaks are reduced from 1300\,Hz to 300\,Hz for the normal hierarchy and 600\,Hz for the inverted hierarchy.
The peaks correspond to the peaks of the luminosity (Fig.~\ref{luminosity.pdf}) and average energy (Figs.~\ref{average_energy.pdf} and \ref{rms_energy.pdf}) of electron neutrinos. 
The ES rates fall to 100\,Hz at around 20 milliseconds and gradually decrease to 0.3\,Hz at 20 seconds.
In the right panel of Fig.~\ref{eventrate_ver3} there are no differences in either of the IBD or ES rates for neutrino oscillations.
As a result, neutrino oscillations have little impact on the observation except for the peak rate of the ES events. 

\begin{table}[H]
    \begin{tabular}{ccccccc}
    \hline
       & $N_{\rm tot}$ & $N(0\leq t \leq 0.3)$ & $N(0.3 \leq t \leq 1)$ & $N(1 \leq t \leq 10)$ & $N(10 \leq t \leq 20)$ \\\hline
       IBD (No osc)& 1782.6 & 575.6(32.3\%) & 377.8(21.2\%) & 682.0(38.3\%) & 147.1(8.25\%)\\
        ES (No osc) & 89.9 & 21(23.3\%) & 27.6(30.7\%)& 34.1(37.9\%)& 7.22(8.04\%)\\
       IBD (Normal) & 1792.6 & 558.7(31.2\%) & 375.7(21\%)& 706.2(39.4\%)& 152.0(8.48\%)\\
       ES (Normal) & 80.3 & 7.7(9.59\%) & 28.5(35.5\%)& 36.4(45.4\%)& 7.64(9.52\%)\\
       IBD (Inverted) & 1814.9 & 520.9(28.7\%) & 371.1(20.4\%)& 760.0(41.9\%)& 163.0(8.98\%)\\
       ES (Inverted) & 83.4 & 11.9(14.3\%) & 28.1(33.7\%)& 35.9(43\%)& 7.53(9.03\%)\\
       \hline
    \end{tabular}
    \caption{Number of events at SK for a supernova at 10 kpc. For each reaction $N_{\rm tot}$ is the total number of events, $N(t_{\rm min}\leq t \leq t_{\rm max})$ is the number of events in the time interval between $t_{\rm min}$ and $t_{\rm max}$, and the number in the brackets is the ratio relative to $N_{\rm tot}$. \label{numberofevents}}
\end{table}

\begin{figure}[htbp]
    \centering
    \includegraphics[width=9.5cm]{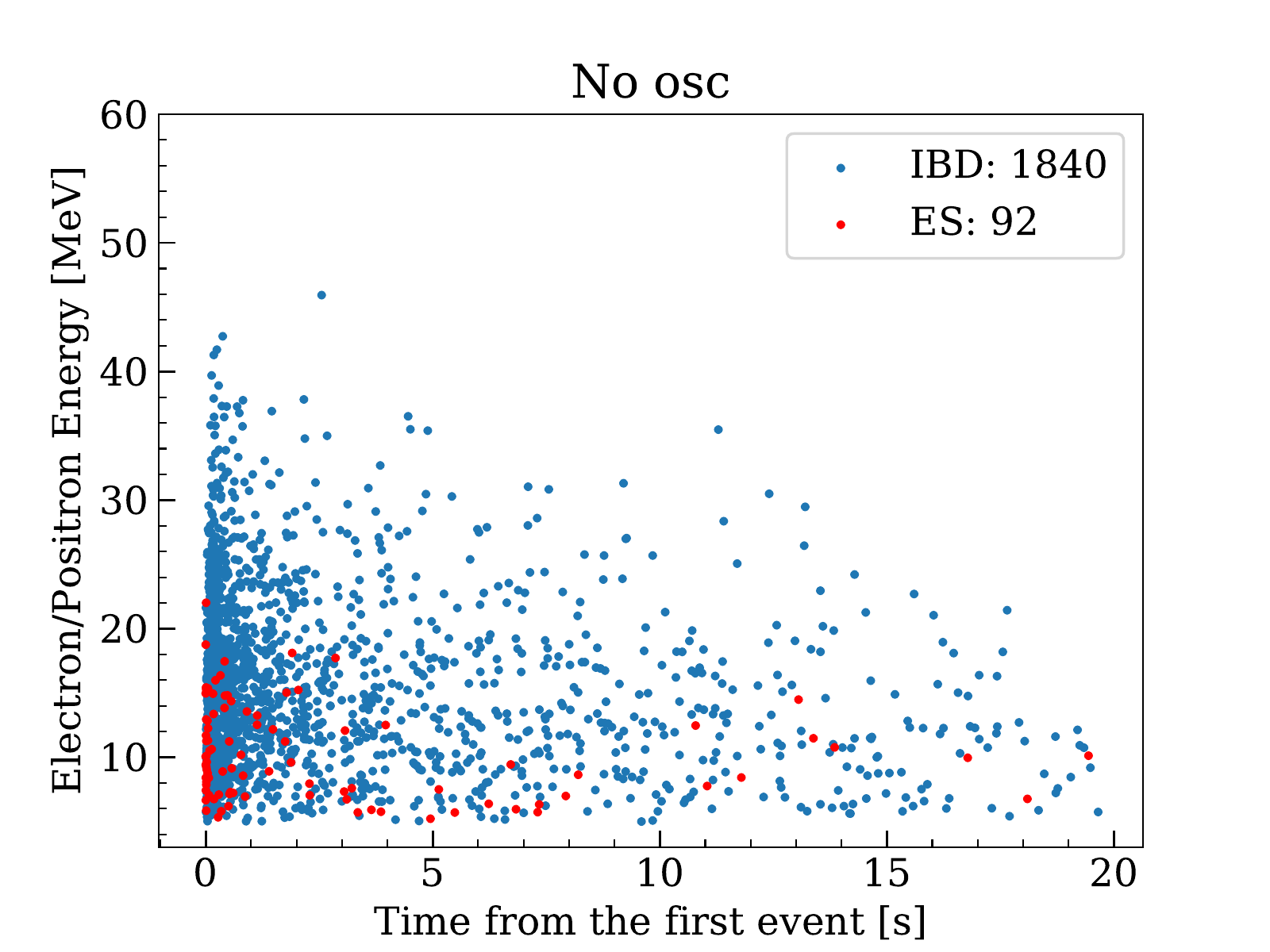}
    \includegraphics[width=9.5cm]{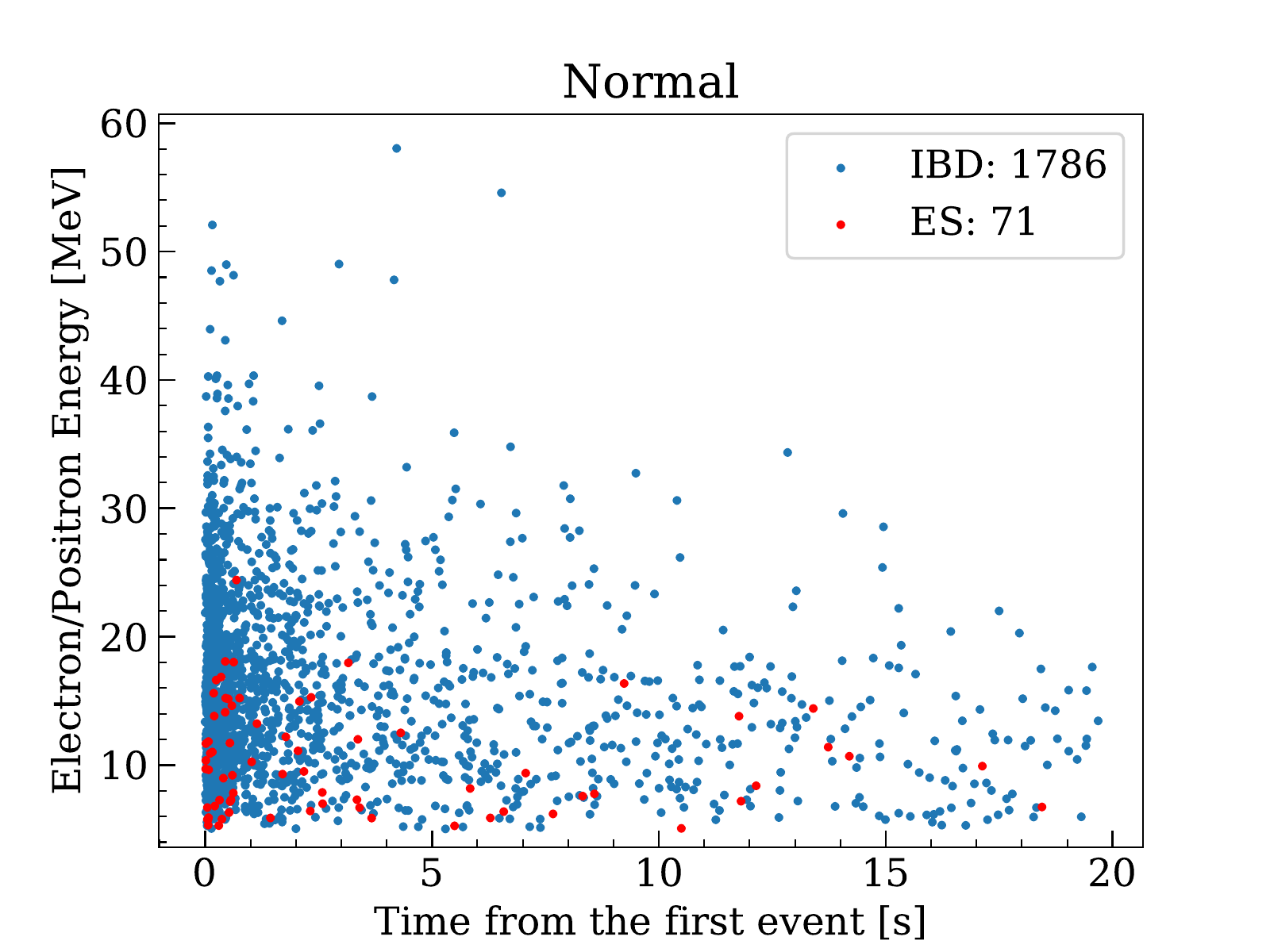}
    \includegraphics[width=9.5cm]{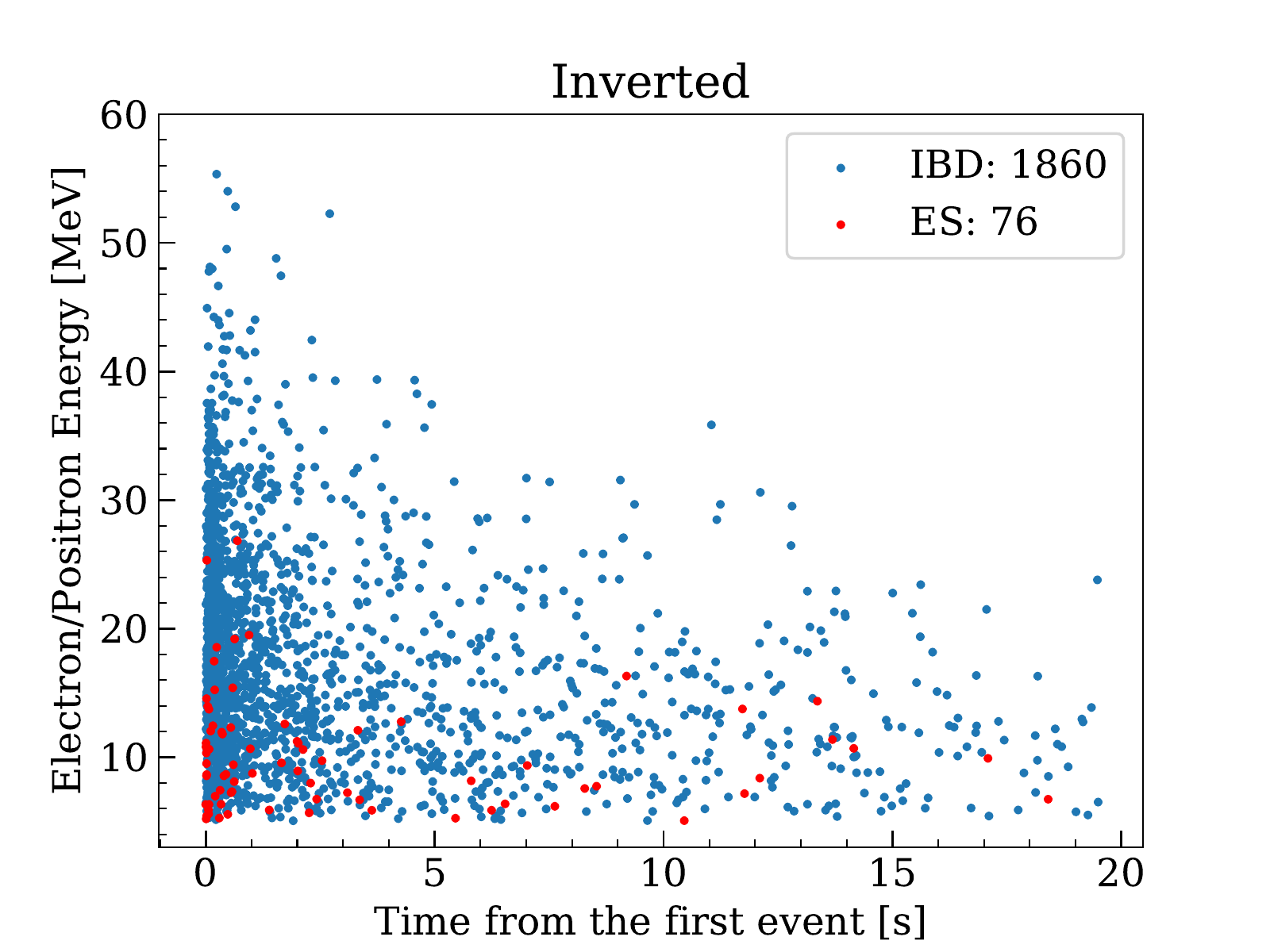}
    \caption{IBD and electron scattering for a supernova at 10 kpc assuming the 32.5 kton inner detector volume of SK. The vertical axis shows electron or positron energy after the neutrino reaction. The top is for no oscillation, the middle is for the normal hierarchy and the bottom is for the inverted hierarchy.\label{event_rate_ibd_es}}
\end{figure}

We show a scatter plot of the event time and lepton energy in Fig. \ref{event_rate_ibd_es}.
As seen from this figure, the fraction of events with energy greater than 30~MeV declines with time as is 
expected from the decline in the average neutrino energy shown in Figs. \ref{average_energy.pdf} and \ref{rms_energy.pdf}.

\begin{figure}[htbp]
    \centering
    \includegraphics[width=10cm]{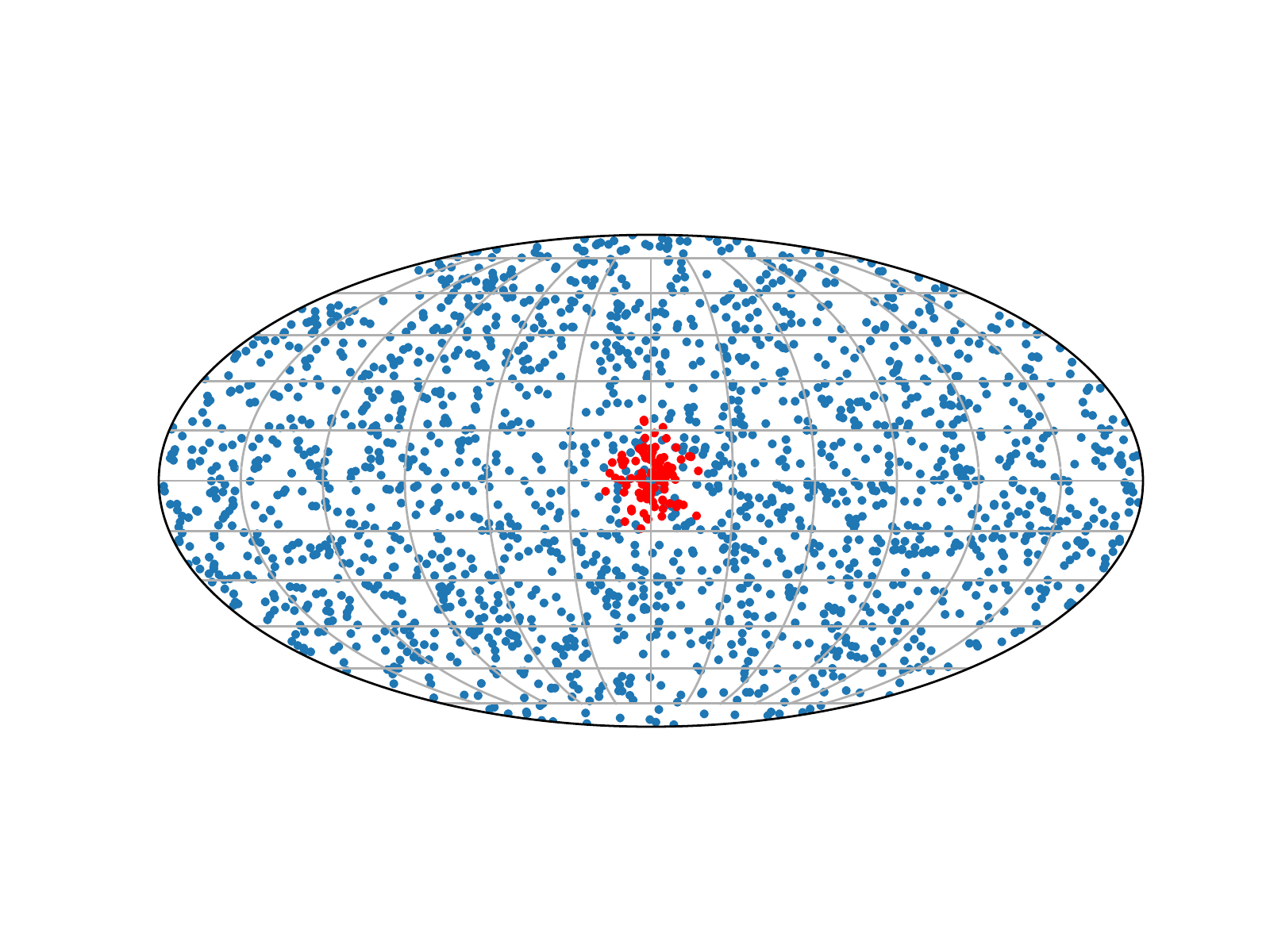}
    \caption{Direction of scattered electrons and positrons, assuming that a supernova happens at the galactic center, 10 kpc from the Earth. The blue dots are IBD events and the red dots are electron scattering events.}
    \label{event_skymap}
\end{figure}

The distribution of true event directions is shown in Fig. \ref{event_skymap}. Here the supernova is assumed to have occurred at the center of the plot and the horizontal (vertical) direction represents event latitude (longitude). 
Blue points are IBD events and the red points are electron scattering events. 
Note that IBD events are distributed throughout the plot, since the outgoing lepton does not preserve the direction of the incoming neutrino, while the electron scattering events concentrate at the center.
As a result, the latter can be used to point back to the supernova. 

\begin{figure}
    \centering
    \includegraphics[width=14cm]{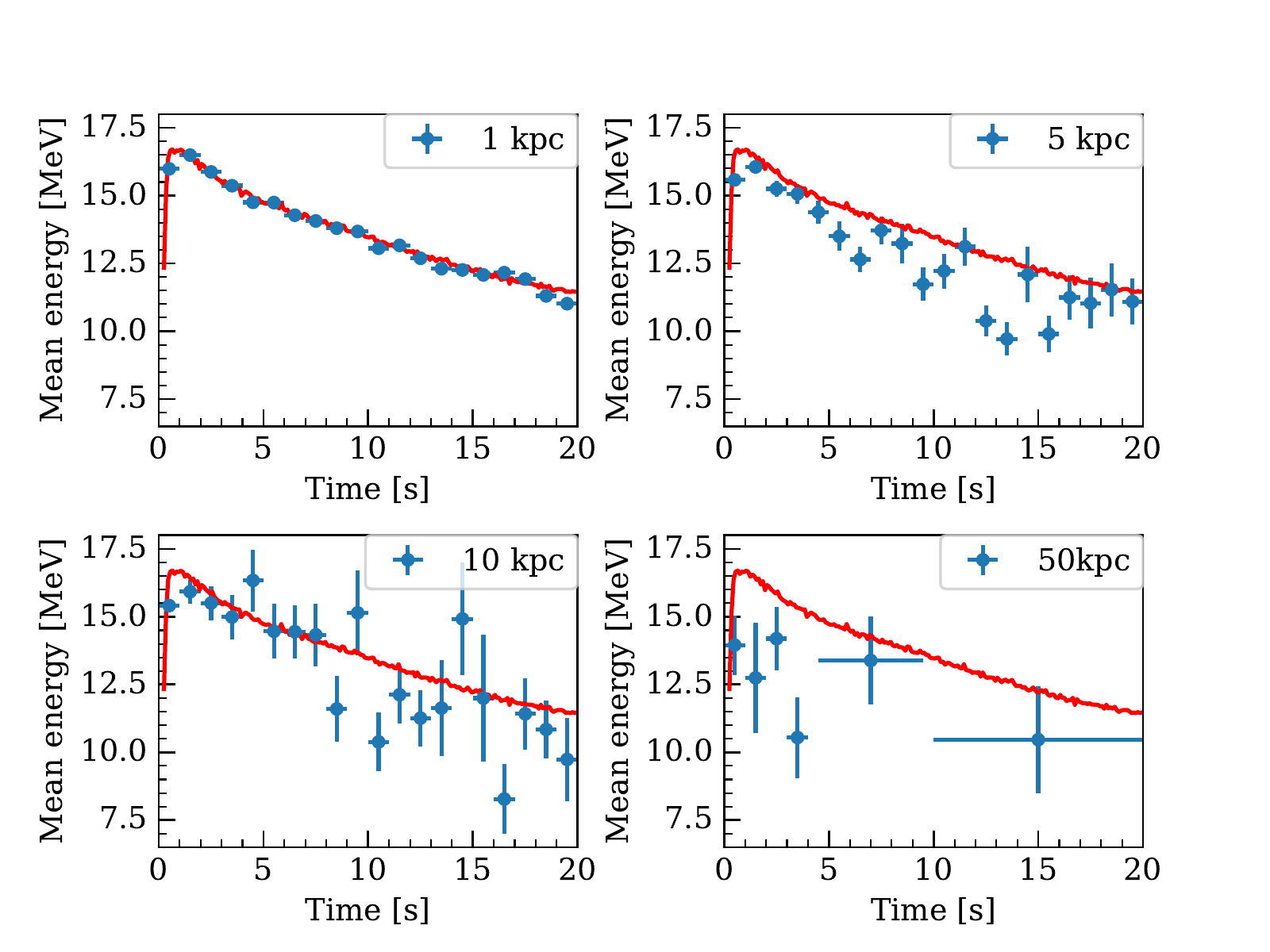}
\caption{The evolution of the mean energy evolution of positrons from  IBD events (blue) for no oscillation. The red curves show the theoretical expectation.  Horizontal bars show the width in time over which the mean energy is calculated. The width of time bin is normally 1 second, though the last 2 bins are 5 seconds and 10 seconds for the 50 kpc model as there are fewer events at late times. There are no events in the last 1 second in this model. The average energies are likely to be below the theoretical values due to the asymmetric shape of the theoretical energy distribution and the limited statistics of an observation at this distance.}
    \label{event_average}
\end{figure}

Figure~\ref{event_average} shows the time evolution of the mean energy of IBD events for no oscillation observed at each assumed supernova distance. 
In this figure the red curves show the theoretical expectation and the blue points show examples from the mock data sets.  
For the latter the width of the time bins is taken to be 1 second, except for the last two bins of the 50 kpc model.
Here the error bars $E_{\rm err}$ are defined as
\begin{equation}
    E_{\rm err} = \sqrt{\dfrac{\dfrac{1}{N_{\rm bin}}\sum^{N_{\rm bin}}_{i=1}(E_i-\bar{E})^2}{N_{\rm bin}}},
\end{equation}
where $N_{\rm bin}$ is the number of events, $E_i$ is the positron energy of the $i$-th event and $\bar{E}$ is the average energy of events in the time bin. 
The mock sample for the supernova at 1 kpc has mean energies that are almost the same as the theoretical curve. 
For the mock samples with from supernovae at 5 kpc and 10 kpc, the difference in the mean energies from the theoretical curve is less than 1 MeV before 5 seconds, but is larger after 5 seconds. 
However, the time evolution of the mean energies reproduces the theoretical curve well. 
For the mock sample at 50 kpc the time evolution does not track the expectation.
Note that here we only consider only statistical uncertainties, which are symmetric, 
while the neutrino energy distribution in each time bin is asymmetric. 
In this sense the mock data distribution does not match the expectation well due to insufficient statistics.
We will consider asymmetric errors including the effect of the shape of the distribution in the future. 

\begin{figure}[htbp]
   \centering
    \begin{tabular}{ccc}
         \begin{minipage}{0.33\hsize}
         \centering
         \includegraphics[width=6.1cm]{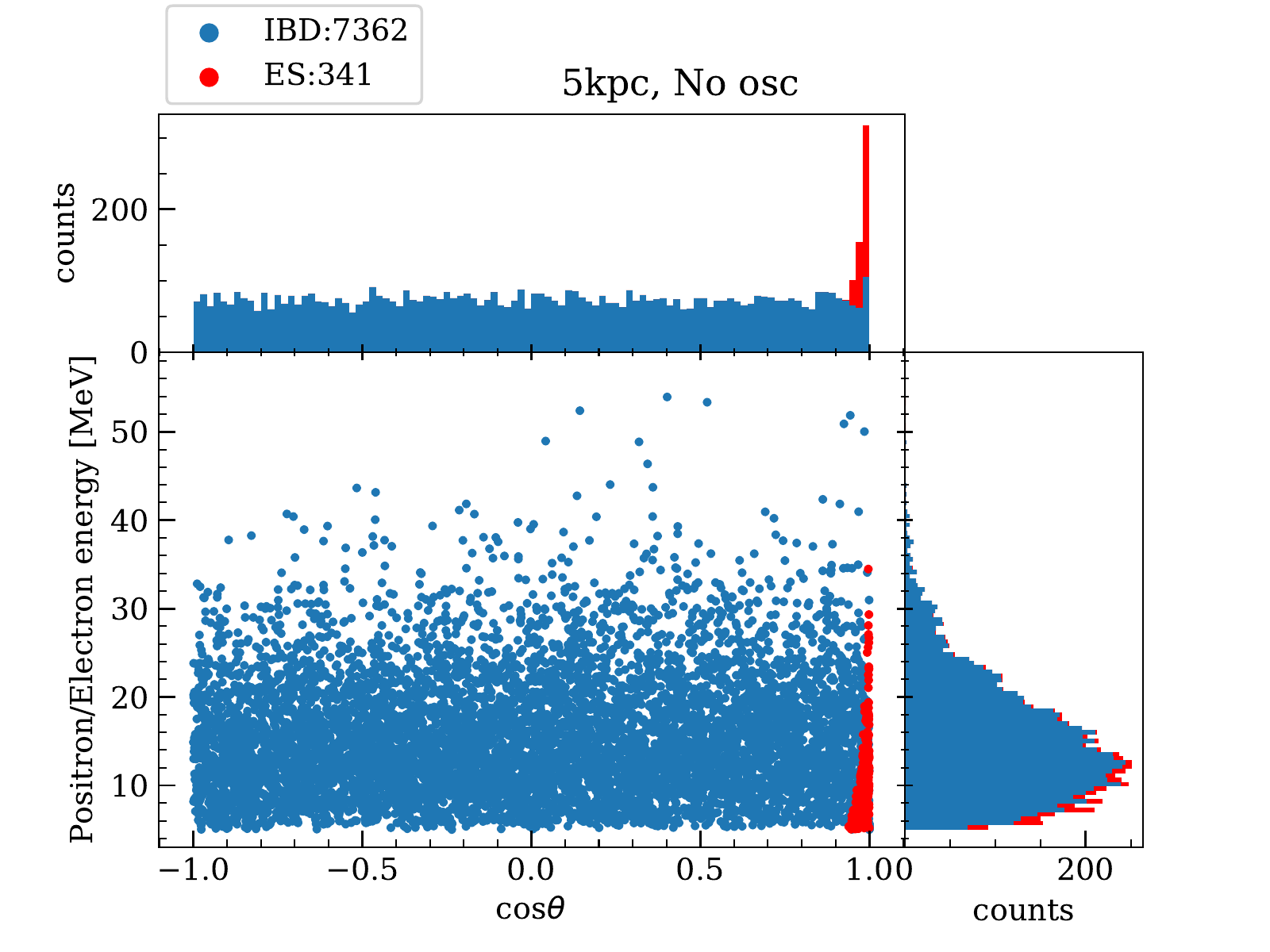}
         \end{minipage}&
         \begin{minipage}{0.33\hsize} 
         \centering
         \includegraphics[width=6.1cm]{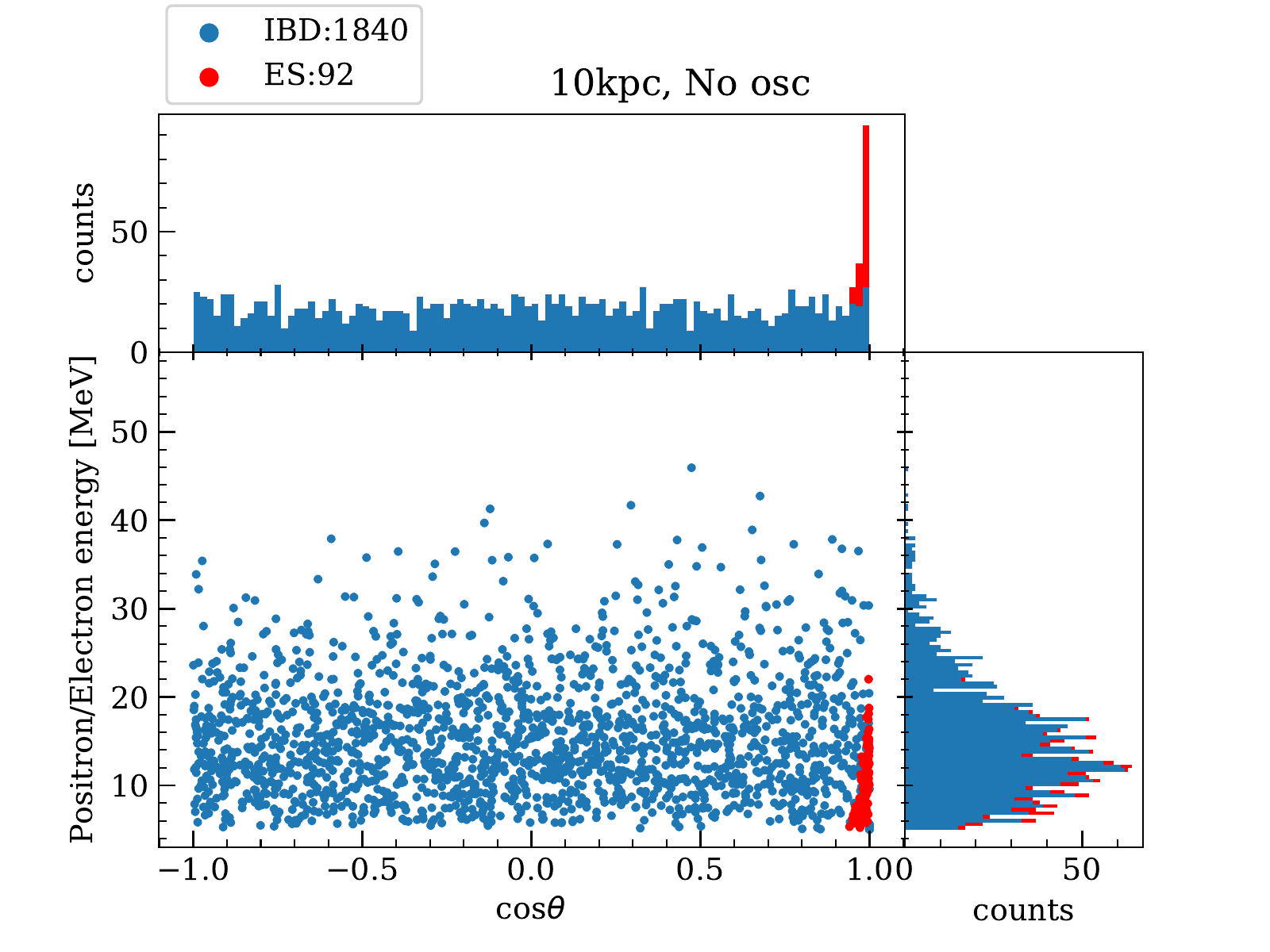}
         \end{minipage}& 
        \begin{minipage}{0.33\hsize}
         \centering
         \includegraphics[width=6.1cm]{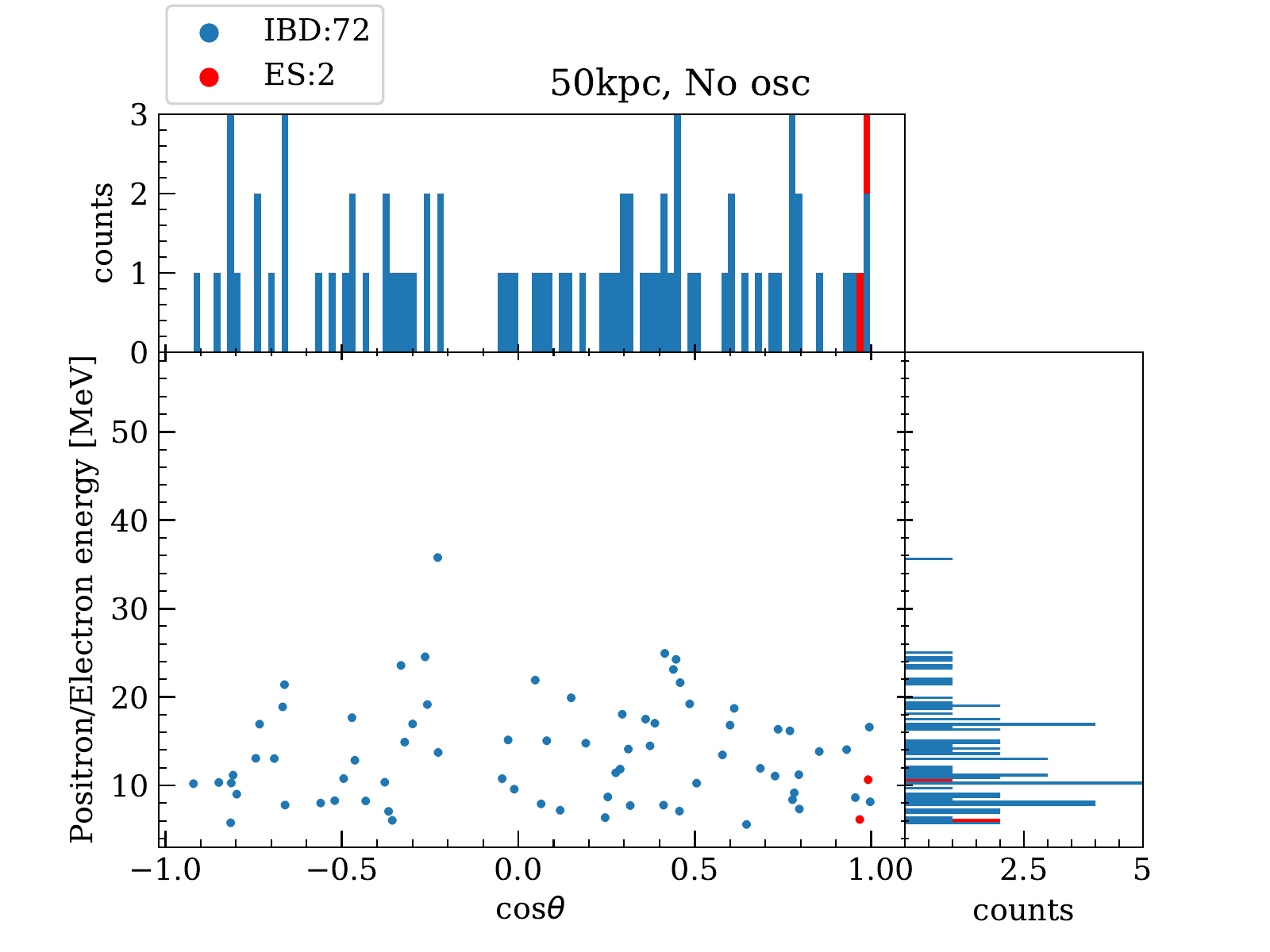}
         \end{minipage}\\
        \begin{minipage}{0.33\hsize}
         \centering
         \includegraphics[width=6.1cm]{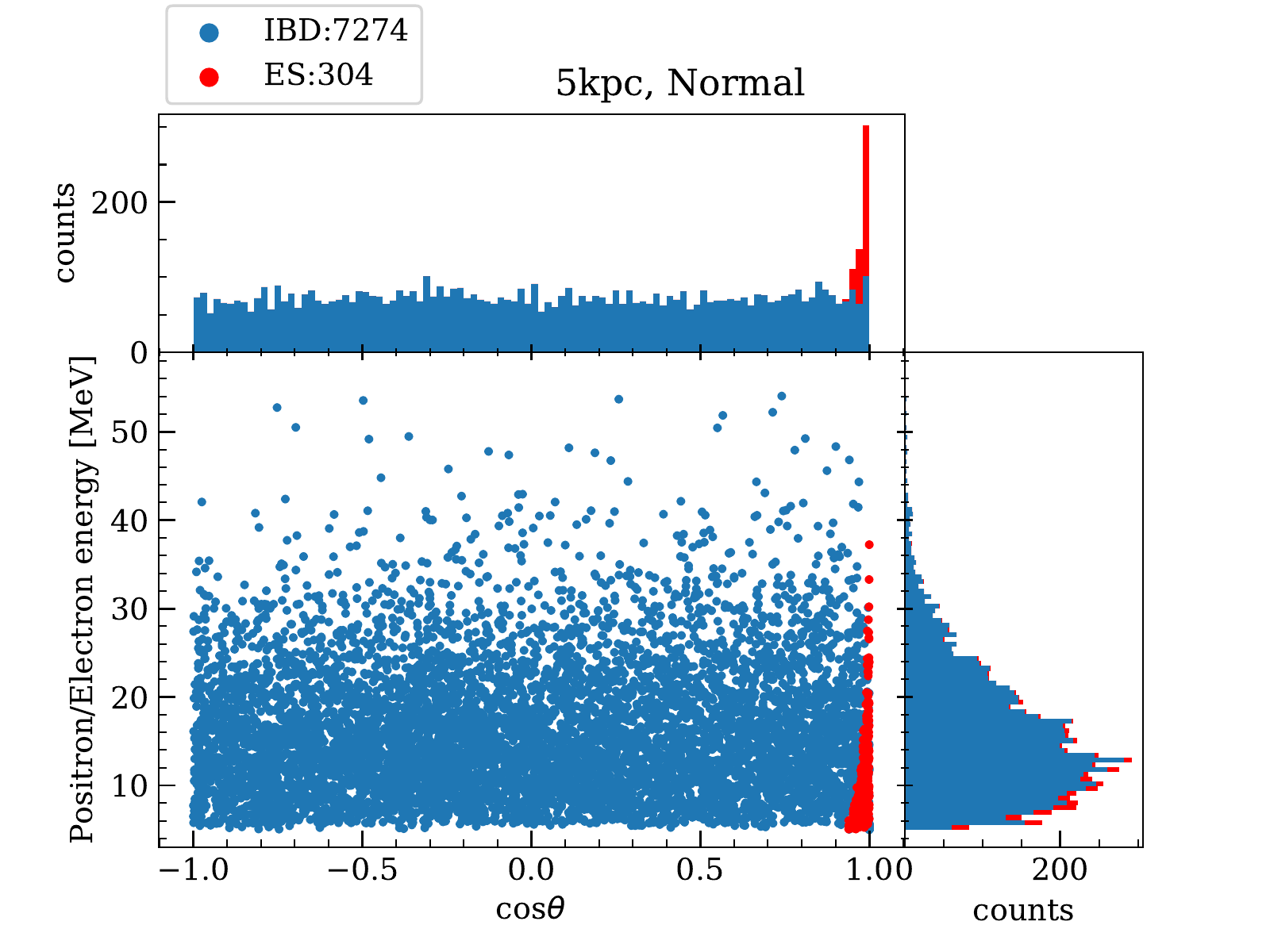}
         \end{minipage}&
         \begin{minipage}{0.33\hsize} 
         \centering
         \includegraphics[width=6.1cm]{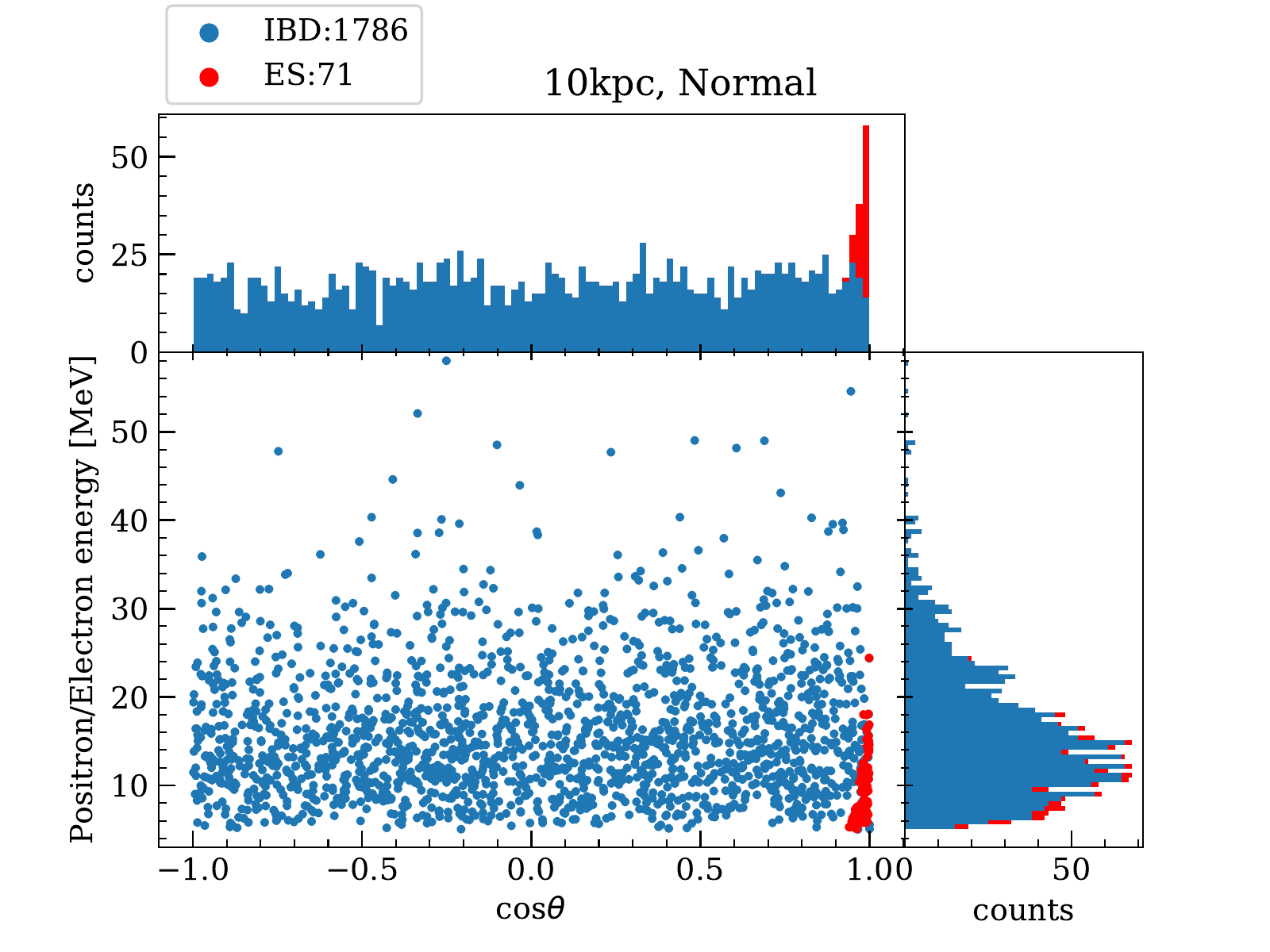}
         \end{minipage}& 
        \begin{minipage}{0.33\hsize}
         \centering
         \includegraphics[width=6.1cm]{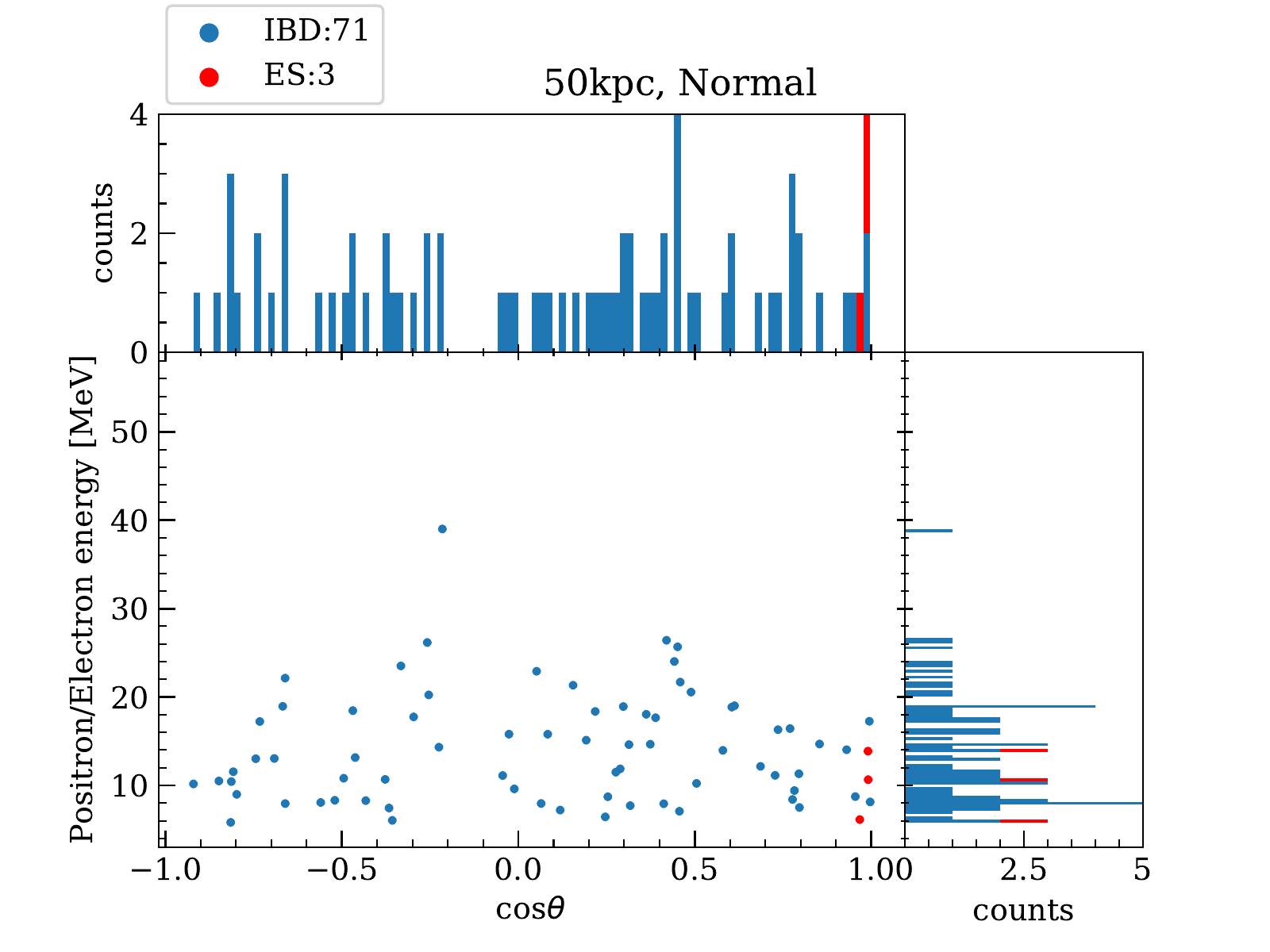}
         \end{minipage}\\
                 \begin{minipage}{0.33\hsize}
         \centering
         \includegraphics[width=6.1cm]{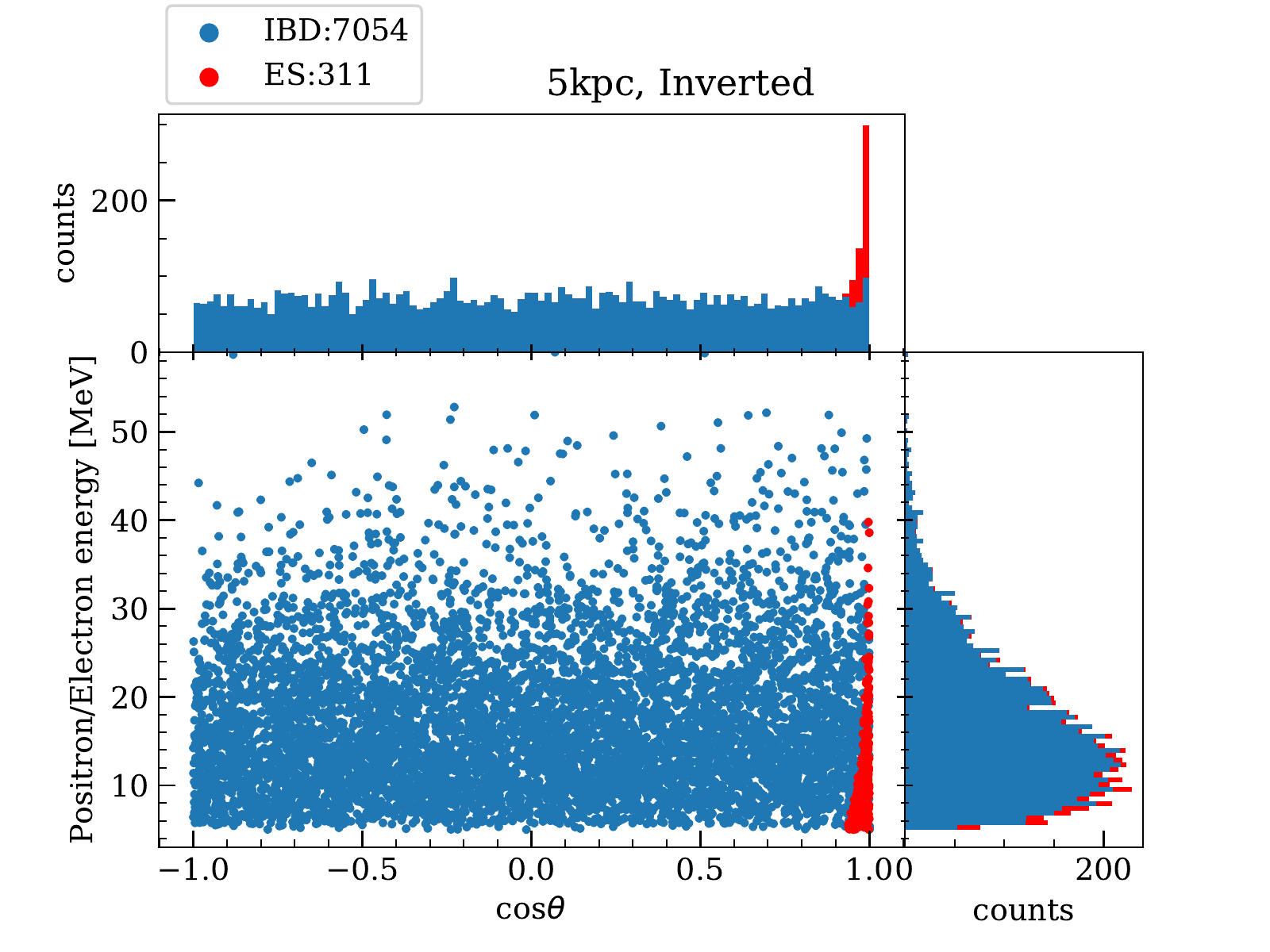}
         \end{minipage}&
         \begin{minipage}{0.33\hsize} 
         \centering
         \includegraphics[width=6.1cm]{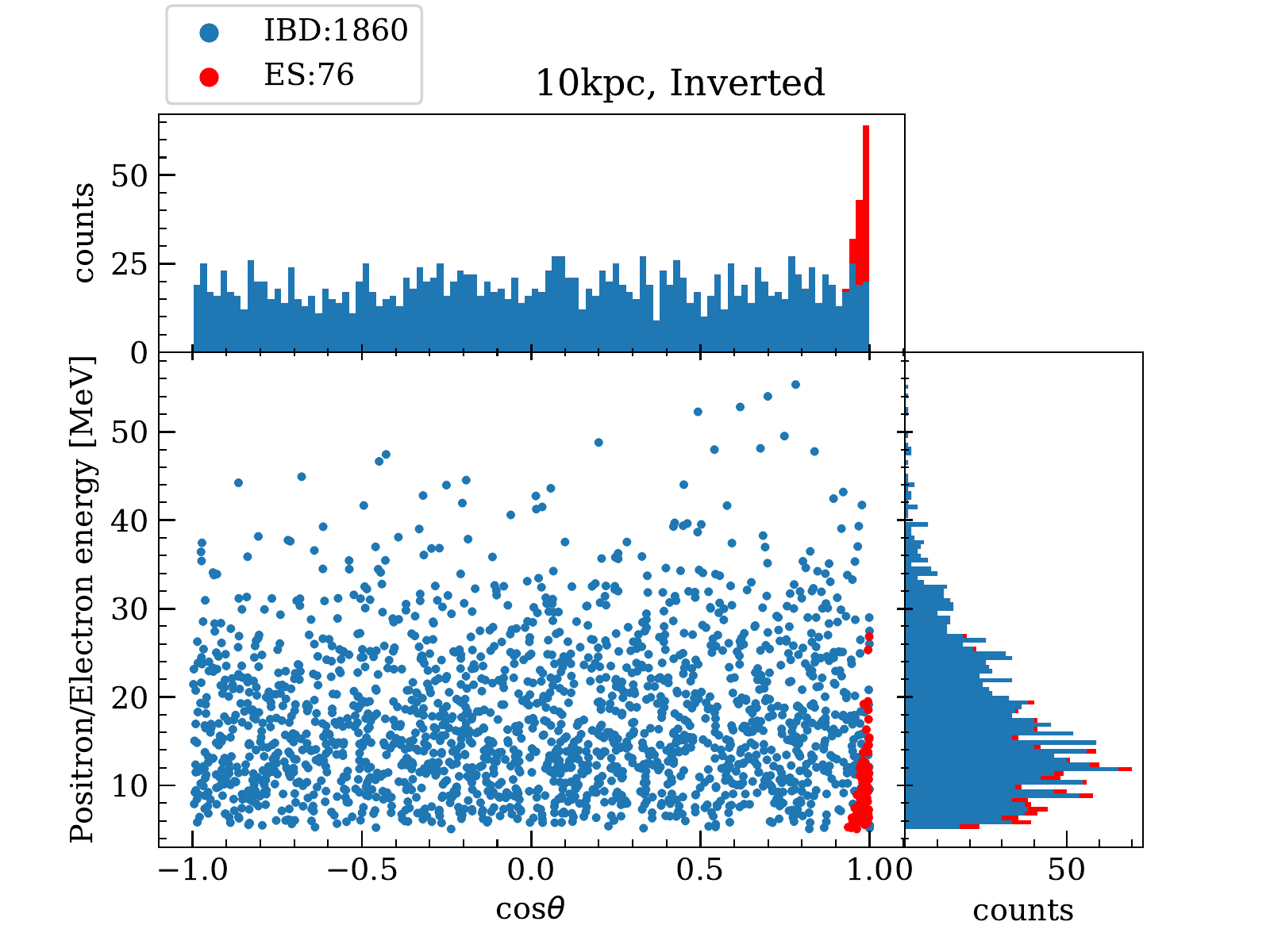}
         \end{minipage}& 
        \begin{minipage}{0.33\hsize}
         \centering
         \includegraphics[width=6.1cm]{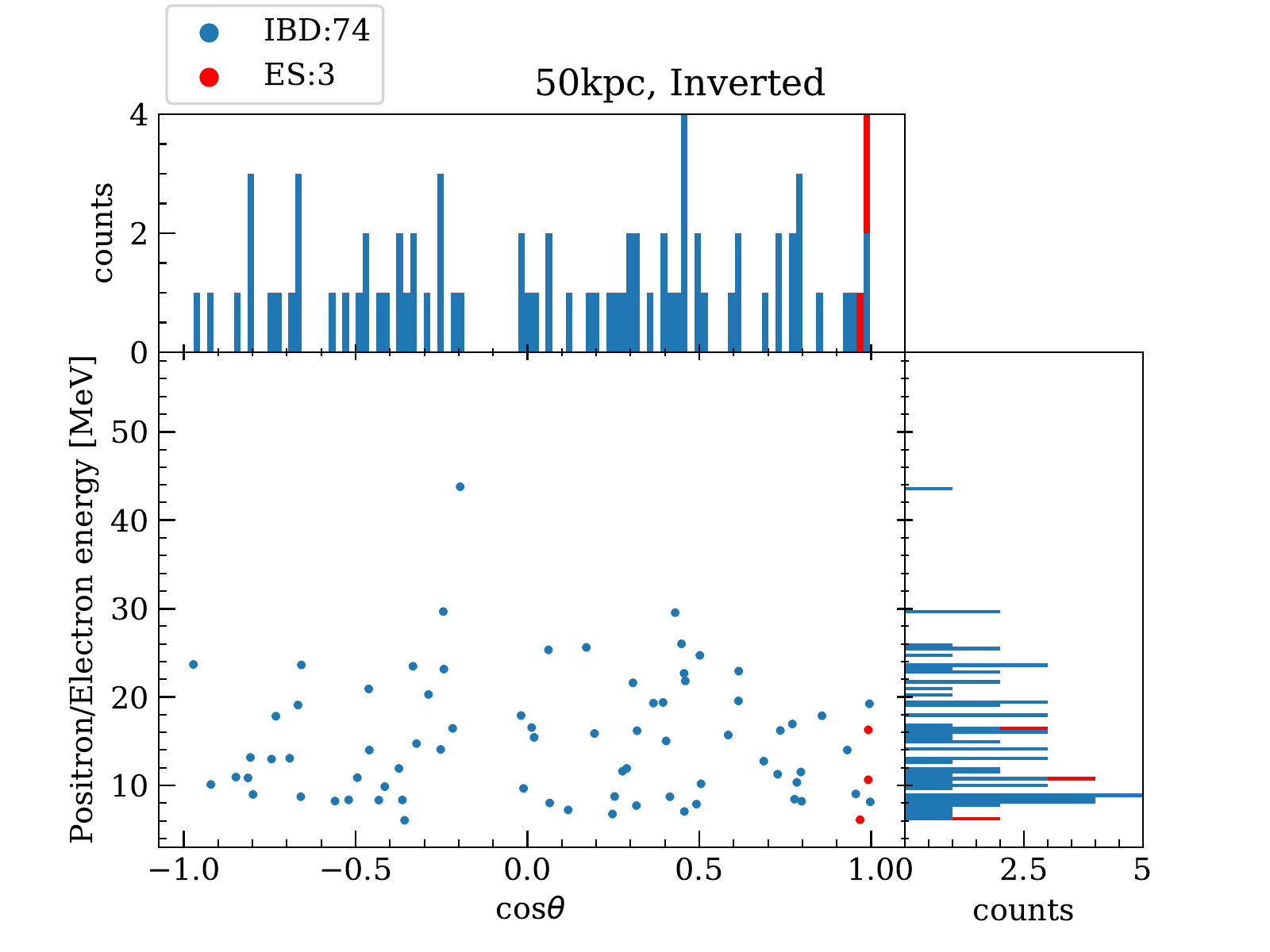}
         \end{minipage}
    \end{tabular}
    
\caption{Distribution of $\cos\theta$ and energy of charged particles from IBD (blue) and electron scattering (red) reactions. Electron scattering events (red) are shown stacked on IBD events (blue). The left, center and right panels are for the models with the supernova distances of 5 kpc, 10 kpc, and 50 kpc (bottom), respectively. The top, middle and bottom panels are for no oscillation, the normal hierarchy and the inverted hierarchy, respectively. The numbers in the legends indicate the total number of events.\label{energy_cos}}
\end{figure}

Figure~\ref{energy_cos} shows
the charged particle energy and $\cos\theta$ of individual events in the mock samples, where $\theta$ is the angle between the neutrino and the final state charged particle. 
The left panel shows the distributions for the model at 5 kpc, the middle is at 10 kpc, which is the distance to the galactic center, and the right is at 50 kpc, which is the distance to the Large Magellanic Cloud. The top panel assumes no oscillations, the middle is the for the normal hierarchy and the bottom is for the inverted hierarchy. The electron scattering events (red) are shown stacked on the  IBD events (blue).

From Fig. \ref{energy_cos} we can see that the total event number decreases with the supernova distance, as expected. 
The IBD events are flat in the $\cos{\theta}$ distribution. On the other hand, electron scattering events have peaks in the $\cos{\theta}$ histograms toward the supernova direction for the mock samples for the supernovae at 5 kpc and 10 kpc. At 50 kpc, the same peak cannot be resolved clearly. 



\subsection{SK-Gd}\label{sec:sk-gd}
\begin{figure}[htbp]
    \centering
    \includegraphics[width=10cm]{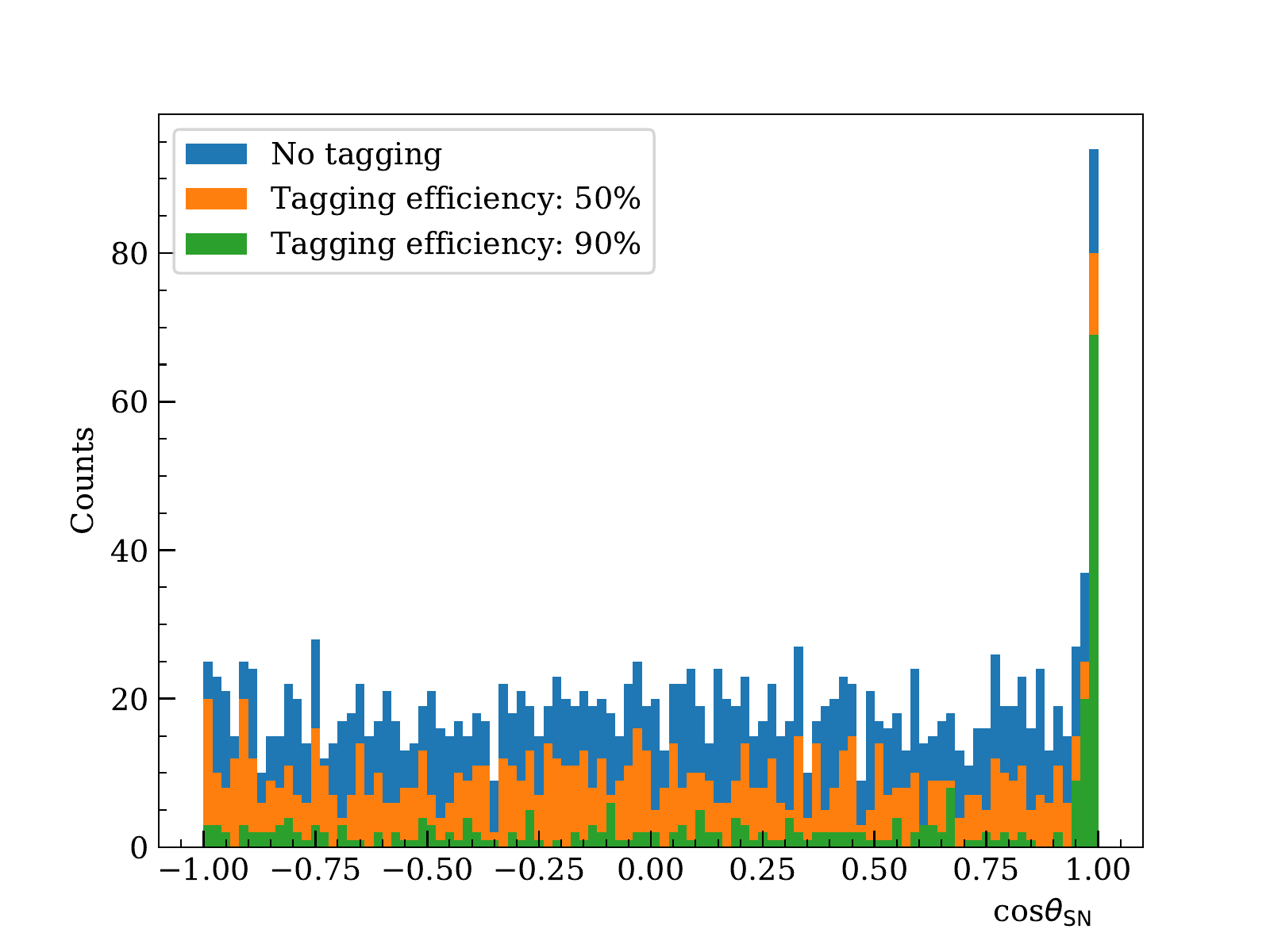}
    \caption{Comparison of angular distributions of IBD and electron scattering events for different neutron tagging efficiencies assuming a supernova at 10 kpc. The blue, orange and green histograms are for no tagging, 50\%, and 90\% tagging efficiency,  respectively.}
    \label{skgd_compare}
\end{figure}

The next stage of SK operations, known as SK-Gd \cite{Beacom_2004}, started in 2020. For the SK-Gd period a gadolinium compound has been dissolved into the pure water to improve the detector's ability to tag neutrons from IBD. 
SK-Gd mainly aims to detect the diffuse supernova neutrino background, however, it  is also expected to improve the ability to determine the location of a supernova burst. 
Indeed, neutron tagging can be used to remove IBD events, leaving only electron scattering events for a precise determination of the supernova direction. 
Figure \ref{skgd_compare} shows the angular distributions of IBD and electron scattering events assuming no neutron tagging, $50\%$, and $90\%$ tagging efficiency for a supernova at 10 kpc. 
It is clear that improved neutron tagging enhances the forward scattering peak. 

\subsection{Comparison with SN 1987A}\label{sec:compare_sn1987a}

\begin{figure}[htbp]
  \centering
  \includegraphics[width=7cm]{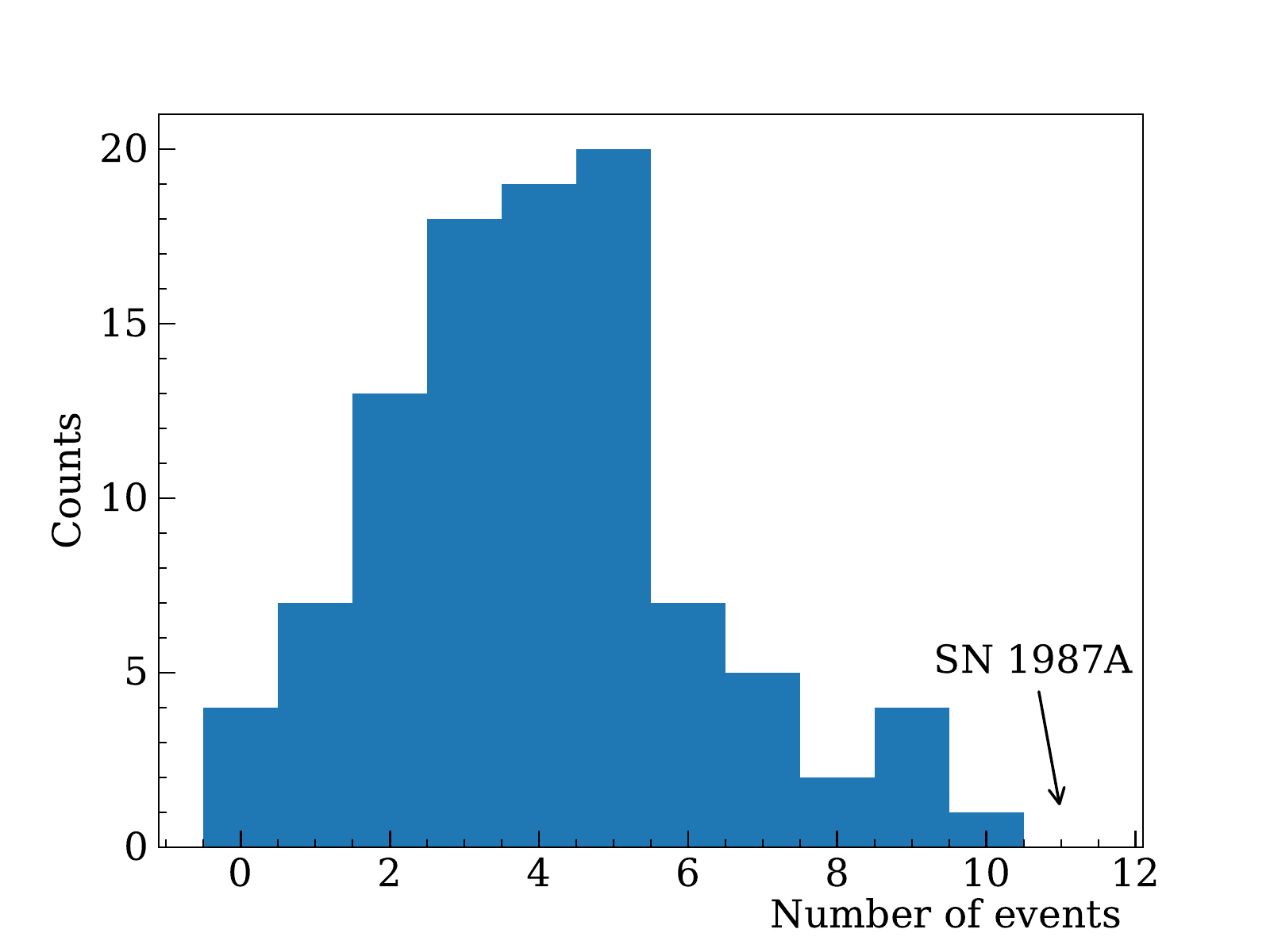}
  \includegraphics[width=7cm]{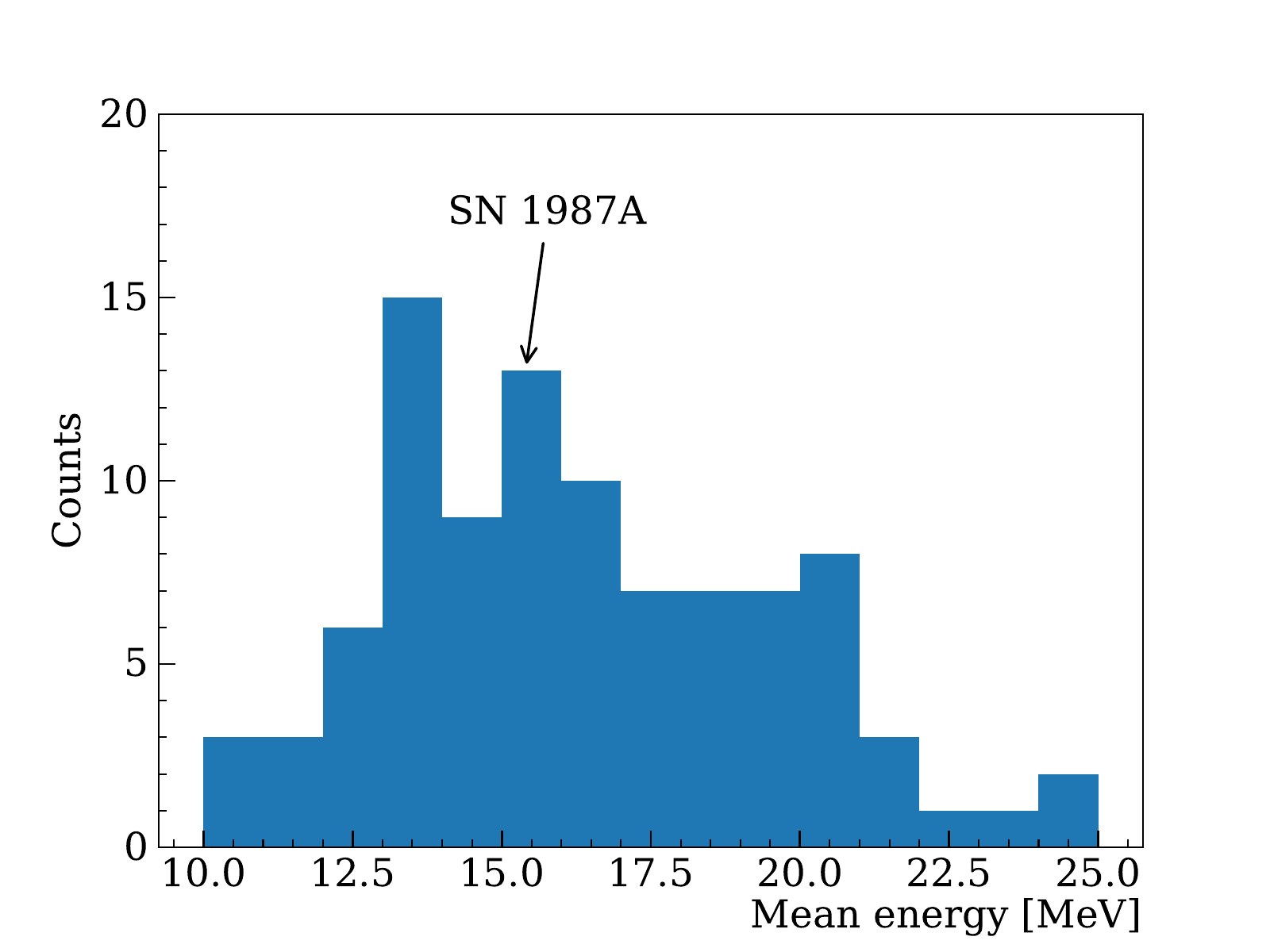}
  \caption{\label{z9_6_ver2_sn1987a_nevent} 
  Results of 100 Monte Carlo simulations that assume a z9.6 supernova happens at 51.4 kpc. Both figures are assumed to be observed with the 2.14 kton target of Kamiokande where neutrino events with the energy lower than 7.5 MeV are not included. The left histogram shows the total number of events and has a mean of four in our model. The right histogram shows for the mean energy of the neutrino events and has an average of 16.4 MeV.
}
\end{figure}

To demonstrate the supernova analyzer in our framework, we compare our model with Kamiokande's observation of SN 1987A's neutrinos  \cite{Hirata:1987hu}. For this purpose we assume that the supernova distance is 51.4 kpc, the detector fiducial volume is 2.14 kton, and the detection threshold is 7.5 MeV in this subsection. We realize 100 Monte Carlo simulations using only IBD interactions.
The left panel of Fig. \ref{z9_6_ver2_sn1987a_nevent} shows a histogram of the number of events obtained in these simulations.
Though the average expected observation in our model is four events, 11 events were observed from SN 1987A. 
The right histogram shows the distribution of mean energies, which can be compared with 15.4 MeV observed at Kamiokande from Hirata et al. (1987) \cite{Hirata:1987hu}. In comparison, the average energy of our simulation is 16.4 MeV.

\begin{figure}[htbp]
  \centering
  \includegraphics[width=10cm]{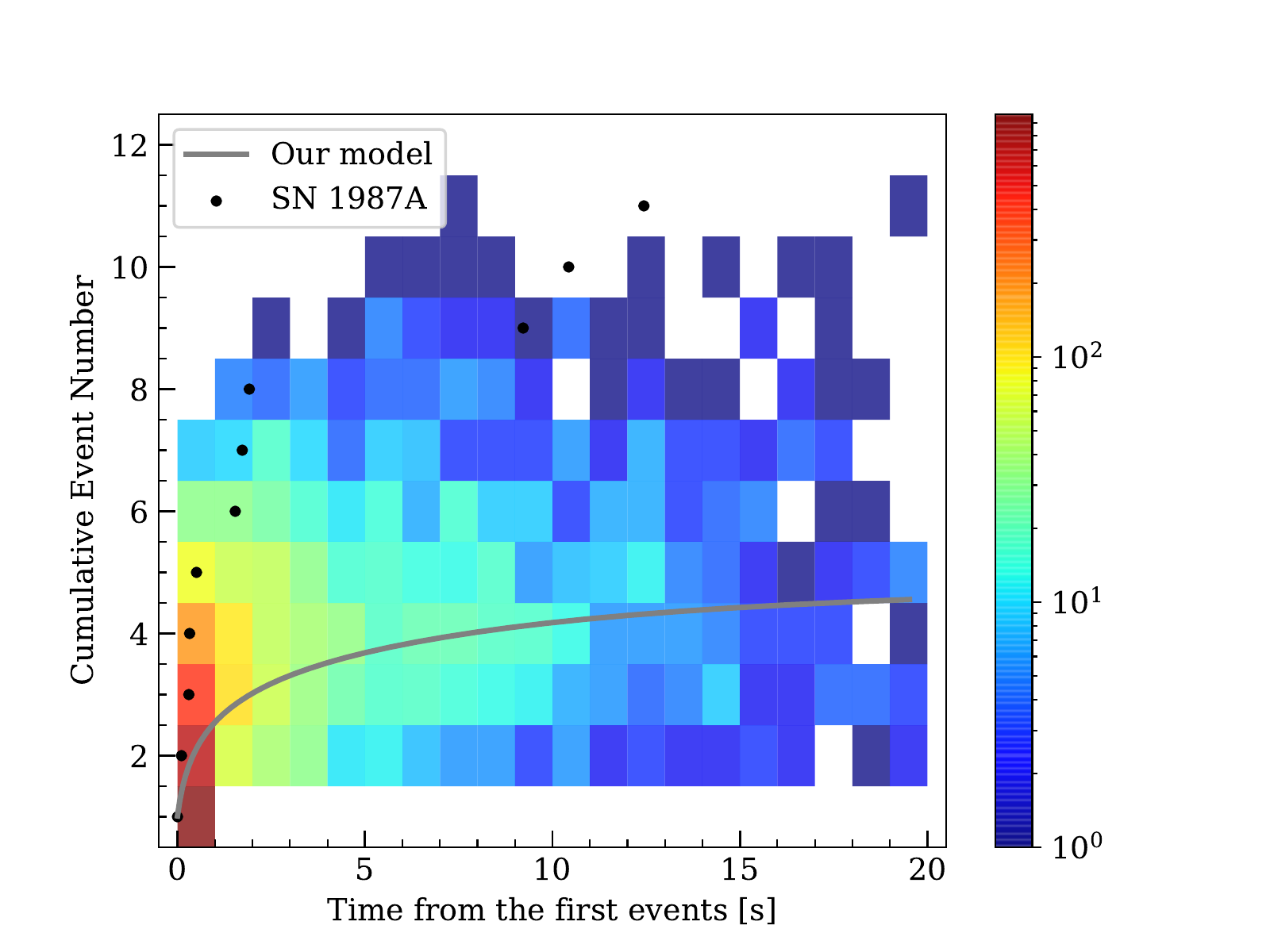}
  \caption{\label{event_time_evo} 
  Comparison of SN 1987A and our model. The figure shows the cumulative number of events over time. The black dots show the time evolution of Kamiokande's observation of SN 1987A. The grey line is average of 1000 cumulative distributions from our simulation. The colored histogram shown the frequency of those cumulative distributions.}
 \end{figure}

\begin{figure}[htbp]
  \centering
  \includegraphics[width=10cm]{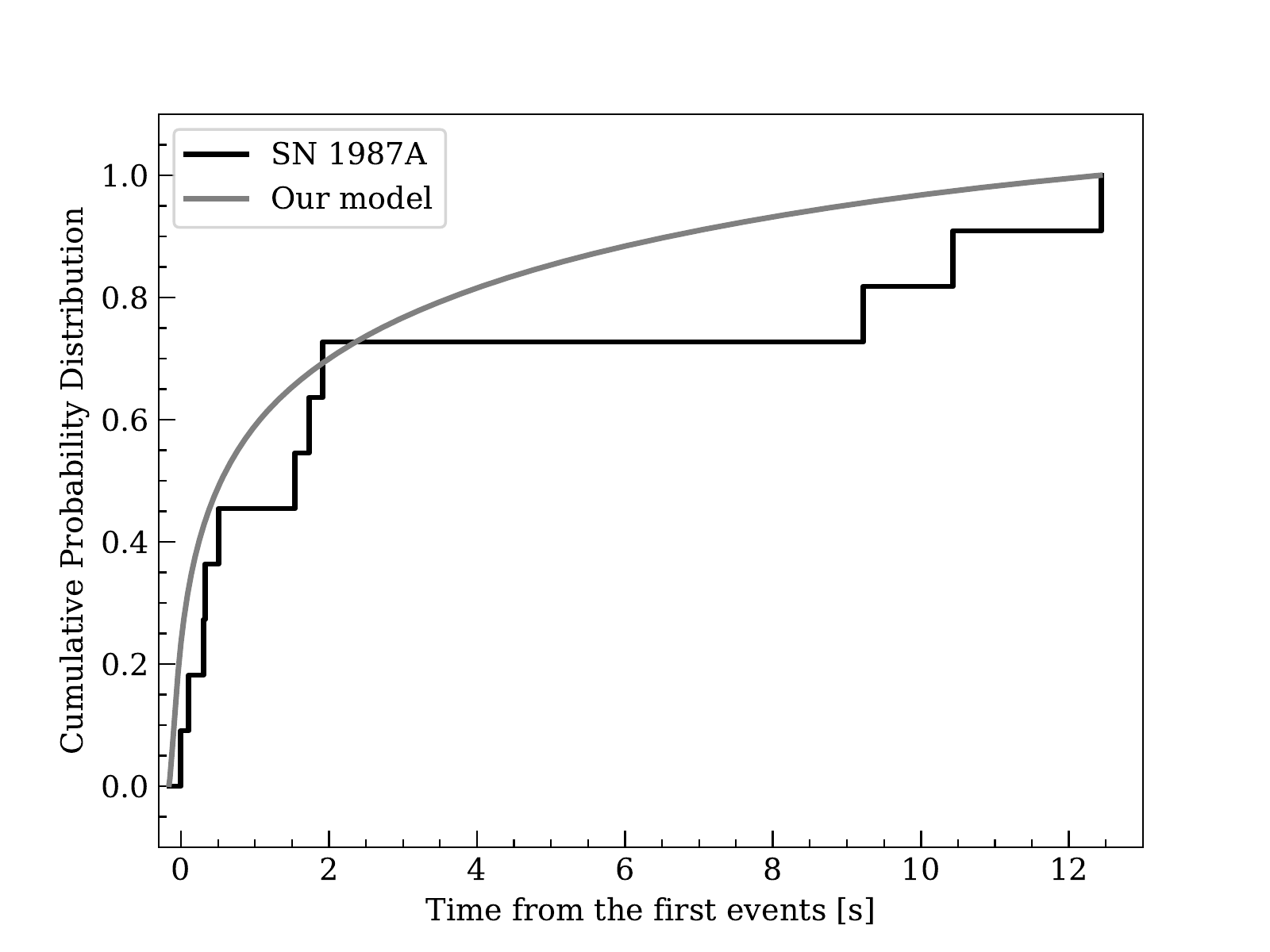}
  \caption{\label{KS_test} KS test using the cumulative event number normalized to 1. The black line is SN 1987A and the grey line is our model.}
 \end{figure}

As described in \S\ref{results} our model corresponds to a weak explosion the number of events on the left side of Fig. \ref{z9_6_ver2_sn1987a_nevent} is within our expectations, though it is lower than the observations of SN 1987A.
With this in mind, we expect the time evolution of the event accumulation to provide a comparison that is less dependent on the total number of events. 
Figure \ref{event_time_evo} shows a comparison of the cumulative number of events (CDF) over time.
Though there are only 11 events from SN 1987A, their accumulation trend is similar to that of our simulation. 
To quantitatively compare with our model we perform a study based on the Kolmogorov-Smirnov test (KS test) in Fig. \ref{KS_test}.
Here the difference in the Kamiokande CDF and the model CDF at 
a time $t$ from the first event is written as $D(t)$.
The maximum distance $|D|_{\rm max}$ between the two is a measure of the compatibility of the two CDFs and is $0.238$. 
For an observation of only 11 events, $|D|_{\rm max} > 0.391$, would indicate an incompatibility of the model and observation at 95\% confidence level. 
Though our model may need more neutrino emission to fully account for the observation of SN 1987A, the two are not inconsistent. 
It is possible that adopting different progenitor simulations may address our model's underestimation of the total number of events.

\section{Summary and discussion}

We have developed a framework to analyze neutrino events for studies of supernova physics and have shown predictions for neutrino emission over 20 seconds. This model follows neutrino emission from the beginning of collapse to the formation of a PNS by making extensions to the GR1D code.
Using this simulation we describe a detector simulator to predict the neutrino signal at Super-Kamiokande.
Mock data samples from our framework provide basic information to compare the simulation with the future observation of supernova burst.  
Assuming a supernova at 10 kpc about 2000 events are expected at SK, which is consistent with the model of Nakazato et al. (2013) \cite{2013ApJS..205....2N}. 
In these simulations we find that electron scattering events from the neutronization burst can be distinguished from the more abundant IBD events, yielding information on the direction to the supernova. 
The gadolinium-loaded SK-Gd detector will improve this discrimination. 
Comparison of our model with the observation of SN 1987A indicates a consistent mean energy but our predicted event number of four is below the observed 11 events.
Work is needed to increase the total flux of neutrinos without affecting the average neutrino energy.
In the next work we will adopt different explosion models using the progenitor generation method of Ref. \cite{Suwa_2016} to further explore the ability of long-time observations to constrain the physics of the burst.


\section*{Acknowledgment}

We thank  Masayuki Nakahata, Evan O'Connor, Hiroki Nagakura and Tomoya Takiwaki for fruitful discussions.
This work was supported by Grants-in-Aid for JSPS Fellows (JP20J14908, JP20J20189), Grants-in-Aid for Scientific Research (JP19K03837, JP19K23435, JP20H00162, JP20H00174, JP20H01904, JP20H01905, JP20K03973) and Grants-in-Aid for Scientific Research on Innovative areas, `A Paradigm Shift by a New Integrated Theory of Star Formation', `Gravitational wave physics and astronomy: Genesis', `Exploration of Particle Physics and Cosmology with Neutrinos', and `Unraveling the history of the universe and matter evolution with underground physics' (JP17H06365, JP18H04586, JP18H05437, JP19H05811, JP18H05535, JP20H04747) from the Ministry of Education, Culture, Sports, Science and Technology (MEXT), Japan.
This work was partially carried out by the joint research program of the Institute for Cosmic Ray Research (ICRR), The University of Tokyo
and
the Particle, Nuclear and Astro Physics Simulation Program (No. 2019-002, 2020-004) of Institute of Particle and Nuclear Studies, High Energy Accelerator Research Organization (KEK).
The numerical computations in this study were partly carried out on XC40 at YITP in Kyoto University and on XC50 at CfCA in NAOJ.
This work was also supported by 
MEXT as Program for Promoting Researches on the Supercomputer Fugaku,
``Toward a unified view of the universe: from large scale structures to planets''.


\bibliographystyle{ptephy}
\bibliography{mori_ptep}
%

\end{document}